\documentclass[sigconf,natbib=true]{acmart}
\usepackage{bm}
\usepackage{subfigure}
\usepackage{multirow}
\usepackage[utf8]{inputenc}
\usepackage{kotex}
\usepackage{booktabs}
\usepackage{hyperref}
\usepackage{multicol}

\AtBeginDocument{%
  \providecommand\BibTeX{{%
    \normalfont B\kern-0.5em{\scshape i\kern-0.25em b}\kern-0.8em\TeX}}}

\setcopyright{acmcopyright}

\acmConference[SIGIR '23]{Proceedings of the 46th International ACM SIGIR Conference on Research and Development in Information Retrieval, July 23--27, 2023, Taipei, Taiwan}{July 23--27,
  2023}{Taipei, Taiwan}
\acmPrice{15.00}
\acmISBN{978-1-4503-XXXX-X/18/06}

\begin{document}

\title{Blurring-Sharpening Process Models for Collaborative Filtering}

\author{Jeongwhan Choi}
\affiliation{%
  \institution{Yonsei University}
  \city{Seoul}
  \country{South Korea}
}
\email{jeongwhan.choi@yonsei.ac.kr}

\author{Seoyoung Hong}
\affiliation{%
  \institution{Yonsei University}
  \city{Seoul}
  \country{South Korea}
}
\email{seoyoungh@yonsei.ac.kr}

\author{Noseong Park}
\affiliation{%
  \institution{Yonsei University}
  \city{Seoul}
  \country{South Korea}
}
\email{noseong@yonsei.ac.kr}

\author{Sung-Bae Cho}
\affiliation{%
  \institution{Yonsei University}
  \city{Seoul}
  \country{South Korea}
}
\email{sbcho@yonsei.ac.kr}

\renewcommand{\shortauthors}{Choi, et al.}

\begin{abstract}
Collaborative filtering is one of the most fundamental topics for recommender systems. Various methods have been proposed for collaborative filtering, ranging from matrix factorization to graph convolutional methods. Being inspired by recent successes of graph filtering-based methods and score-based generative models (SGMs), we present a novel concept of blurring-sharpening process model (BSPM). SGMs and BSPMs share the same processing philosophy that new information can be discovered (e.g., new images are generated in the case of SGMs) while original information is first perturbed and then recovered to its original form. However, SGMs and our BSPMs deal with different types of information, and their optimal perturbation and recovery processes have fundamental discrepancies. Therefore, our BSPMs have different forms from SGMs. In addition, our concept not only theoretically subsumes many existing collaborative filtering models but also outperforms them in terms of Recall and NDCG in the three benchmark datasets, Gowalla, Yelp2018, and Amazon-book. In addition, the processing time of our method is comparable to other fast baselines. Our proposed concept has much potential in the future to be enhanced by designing better blurring (i.e., perturbation) and sharpening (i.e., recovery) processes than what we use in this paper.
\end{abstract}



\maketitle

\section{Introduction}
Recommender systems are one representative topic of information filtering. These days a non-trivial portion of the revenue of many global information technology (IT) companies is from advertising and recommendation. In this regard, recommender systems are of utmost interest in real-world environments. Among various technologies, collaborative filtering (CF) is one of the most popular approaches of recommender systems, and many CF-based methods have been proposed.

In particular, graph convolution-based CF methods currently show state-of-the-art accuracy~\cite{He20LightGCN, Mao21UltraGCN, choi2021ltocf, Shen21GFCF,kong2022hmlet}. They represent user-item interactions as a bipartite graph and apply the graph convolutional technology. Among various graph convolutional operations, they all use relatively simple linear or low-pass filters. Surprisingly, these approaches now beat other classical and deep learning-based methods.

\begin{figure}[t]
    \centering
    \subfigure[Score-based generative models (SGMs) use two stochastic processes, one for the forward perturbation and the other for the backward recovery. Since the recovery process is stochastic, it does not typically converge to the original sample $\bm{x}(0)$ but to another similar sample. After training, only the recovery process is used to generate fake samples from random noisy vectors $\bm{x}(T) \sim \mathcal{N}(\bm\mu, \bm\sigma)$.]{\includegraphics[width=\columnwidth]{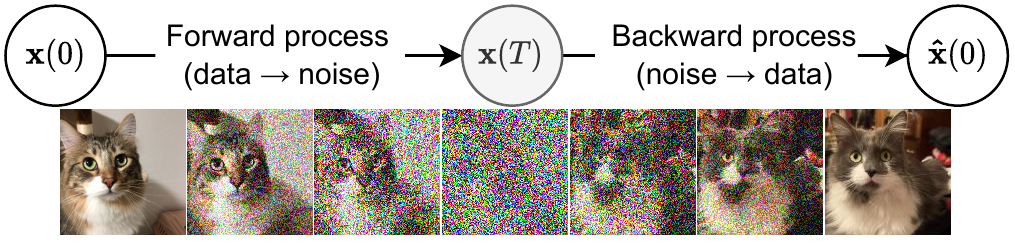}}
    \subfigure[Our blurring-sharpening process models (BSPMs) use two deterministic blurring and sharpening processes. Unlike SGMs trained with many images, our BSPMs process only one interaction matrix and therefore, we use the deterministic processes.]{\includegraphics[width=\columnwidth]{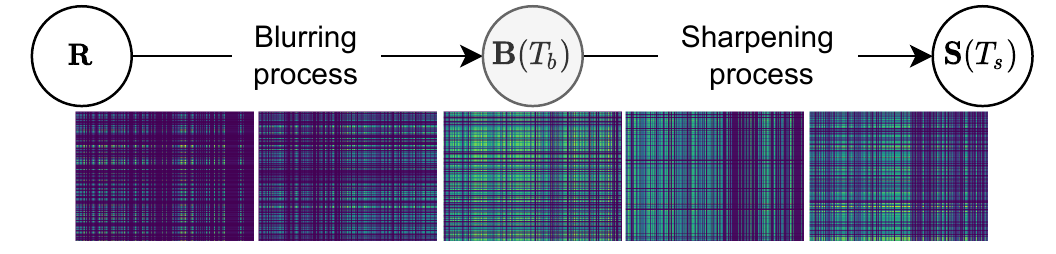}}
    \caption{The comparison between SGMs and our proposed BSPMs. SGMs, a recently proposed paradigm for deep generative task, outperform generative adversarial networks (GANs), variational autoencoders (VAEs), and many other generative models.}
    \label{fig:sgm}
\end{figure}

\begin{table}[t]
    \small
    \setlength\tabcolsep{2pt}
    \caption{Comparison of existing methods. Our BSPMs not only combine the blurring and the sharpening processes but also interpret them in a continuous time domain.}\label{tbl:cmp1}
    \begin{tabular}{ccc}
    \toprule
    Model & Blurring & Sharpening \\ \midrule
    LightGCN & Discrete with heat equation & X \\ 
    LT-OCF & Continuous with heat equation & X \\ 
    GF-CF & Discrete with low pass \& ideal filters & X \\  \midrule
    BSPM & Continuous with various filters & Continuous with a filter \\ \bottomrule
    \end{tabular}
\end{table}

In this paper, we propose a novel paradigm of \textit{\textbf{B}}lurring-\textit{\textbf{S}}harpening \textit{\textbf{P}}rocess \textit{\textbf{M}}odel (BSPM) for CF. Our blurring and sharpening processes are formulated as differential equations --- we will also show that some of the existing graph convolution-based CF methods are special cases of our model.

Our method is greatly inspired by i) score-based generative models (SGMs~\cite{song2021scorebased,song2021maximum,song2020improved}) which are considered as state-of-the-art methods for deep generative tasks, and ii) GF-CF and its following work~\cite{Shen21GFCF,Peng2022Less,Peng2022SVDGCN,hu2022mgdcf,Xia2022FIRE,liu2022parameterfree} which are simple and computationally efficient but show state-of-the-art accuracy. GF-CF does not learn embedding vectors for users/items but directly processes the user-item interaction matrix to derive unknown user-item interactions. Although GF-CF has shown encouraging results, we found that our proposed \emph{perturbation-recovery paradigm}, called BSPM, can significantly outperform it. Similar successes were already made for image generation. For instance, SGMs show the state-of-the-art quality in the domain of image generation. In SGMs, specific types of stochastic differential equations (SDEs) are adopted to describe the forward and the backward processes (cf. Fig.~\ref{fig:sgm} (a)) --- the backward process is considered as a generative model.



Our overall model design has a perturbation-recovery architecture, i.e., the blurring process corrupts (or perturbs) original information in the user-item interaction matrix, and the sharpening process tries to recover the original information in conjunction with promising additional information (cf. Fig.~\ref{fig:sgm} (b)). We apply the blurring and the sharpening processes directly to the interaction matrix in a continuous-time manner whereas existing methods, such as GF-CF, apply certain blurring filters to the matrix in a discrete-time manner (cf. Table~\ref{tbl:cmp1}). To our knowledge, we are the first proposing the blurring-sharpening process paradigm for CF.

Therefore, the key in our model is how to define the blurring and the sharpening processes. Both of them are written as ordinary differential equations (ODEs) in our case (cf. Eqs.~\eqref{eq:blur} and~\eqref{eq:sharp}). We customize various well-known blurring and sharpening functions proposed in various domains different from CF.

After defining our blurring and sharpening processes, we design two variants of BSPM: BSPM-LM and BSPM-EM. These variants differ from each other in how to connect the blurring and the sharpening processes. We then show that some popular existing methods are special cases of our method.

We conduct experiments with 3 benchmark datasets and 43 baselines. Surprisingly, our method beats all existing popular CF algorithms by large margins. There are no existing methods that are comparable to our method in all datasets.


Moreover, our proposed model can also be properly understood from the perspective of classical graph convolutional processing. Therefore, we emphasize that our proposed model has strong theoretical grounds, and it is not by chance that our model marks the best accuracy. Our contributions can be summarized as follows:
\begin{enumerate}
    \item There are two research trends, which inspire us: i) GF-CF and its following work, which are some of the state-of-the-art methods for collaborative filtering, directly process the user-item interaction matrix to reveal unknown user-item interactions without learning embedding vectors, and ii) SGMs adopt the perturbation-recovery paradigm to generate fake images.
    \item We design a perturbation-recovery concept, called \textit{\textbf{B}}lurring-\textit{\textbf{S}}harpening \textit{\textbf{P}}rocess \textit{\textbf{M}}odel (BSPM).
    \item Our BSPMs directly perturb (blur) the user-item interaction matrix, and recover (sharpen) the blurred matrix to derive unknown user-item interactions.
    \item Our method outperforms all existing 43 popular CF methods in the three benchmark datasets. 
    \item To our knowledge, we are the first adopting the perturbation-recovery paradigm for CF. Therefore, one can consider that we propose a new paradigm for CF, and we think that it has much potential in the future by discovering better perturbation and recovery processes than ours.
\end{enumerate}

\section{Preliminaries \& Related Work}
In this section, we review related work and preliminary knowledge: collaborative filter (CF), score-based generative models (SGMs), and ordinary differential equations (ODEs).

\subsection{Collaborative Filtering}
Let $\bm{R} \in \{0,1\}^{|\mathcal{U}| \times |\mathcal{V}|}$, where $\mathcal{U}$ is a set of users and $\mathcal{V}$ is a set of items, be an interaction matrix. $\bm{R}_{u,v}$ is 1 \textit{iff} an interaction $(u,v)$ is observed in data, or otherwise 0. We also define the normalized interaction matrix as $\tilde{\bm{R}} = \bm{U}^{-\frac{1}{2}} \bm{R} \bm{V}^{-\frac{1}{2}}$, where $\bm{U} = Diag(\bm{R}  \bm{1})$, $\bm{V} = Diag(\bm{1}^{\mathsf{T}} \bm{R})$, $\bm{1}$ means a column vector of ones, and $^{\mathsf{T}}$ means transpose. We also define the normalized item-item adjacency matrix as $\tilde{\bm{P}} = \tilde{\bm{R}}^{\mathsf{T}}\tilde{\bm{R}}$.

\paragraph{Matrix Factorization-based Methods} 
The most common CF paradigm is to learn latent features (also known as embedding vectors) to represent users and items. The dot product of user and item embedding vectors $\bm{e}_u^T\bm{e}_i$ approximates user $u$’s rating on item $i$, which is denoted by $r_{ui}$. Earlier CF models focused on low-rank matrix factorization (MF)~\cite{koren2009MF}, which aims to approximate the interaction matrix $\bm{R}_{u,v}$. Singular value decomposition (SVD) was initially proposed to learn the feature matrices, followed by many other MF methods~\cite{Mnih2007PMF,rendle2009BPR,Rao2015GRMF,He2017NCF,Yi2018LRML,yang2018hop,chen2020ENMF}. NCF~\cite{He2017NCF} replaces the dot product with a similarity learned with a multi-layer perceptron (MLP). GRMF~\cite{Rao2015GRMF} smoothes MF through adding a graph Laplacian regularizer. HOP-Rec~\cite{yang2018hop} proposes a unified and efficient method that incorporates both MF and graph-based models for CF. ENMF~\cite{chen2020ENMF} proposes a simple neural MF method without a negative sampling strategy and uses an MSE loss function. MF-CCL~\cite{mao2021simplex} proposes a neural MF model trained with a cosine contrastive loss, and shows that it is superior to existing loss functions.



\paragraph{Graph-based Methods} From the perspective of the user-item interaction graph, the individual interaction history is equivalent to the first-order connectivity of the user. Thus, a natural extension is to mine the higher-order connectivity from the user-item graph structure. For example, the second-order connectivity of a user consists of similar users who have co-interacted with the same items. Fortunately, with the development and success of graph convolutional networks (GCNs) for modeling graph structure data in various machine learning areas, it recently became popular to adopt GCNs for CF~\cite{Wang19NGCF,chen20LRGCCF,He20LightGCN,Rianne2017GCMC,Rex2018pinsage,choi2021ltocf,Shen21GFCF,sun2020NIA-GCN,Mao21UltraGCN}.

GC-MC~\cite{Rianne2017GCMC} is the first work using GCNs for recommendations, which is a graph-based auto-encoder framework for explicit matrix completion. PinSage~\cite{Rex2018pinsage} first applies GCNs on web-scale recommender systems and proposes the combination of efficient random walks and GCNs. NGCF~\cite{Wang19NGCF} then proposes an interaction encoder to capture the collaboration signal among users and items, using non-linear activations and transformation matrices. NIA-GCN~\cite{sun2020NIA-GCN} explicitly models the relational information between neighbor nodes and exploits the heterogeneous nature of the user-item bipartite graph.
Graph-based recommendation models have achieved remarkable results, but their efficiency remains unsatisfactory when confronted with large-scale recommendation scenarios. Therefore, improving the efficiency of graph-based methods while leaving high performance for recommendations has become a popular research question. Inspired by a simplified GCN (SGC)~\cite{Wu2019SGC}, LightGCN~\cite{He20LightGCN} outperforms NGCF~\cite{Wang19NGCF} by removing the non-linear activation and feature transformation to improve both accuracy and efficiency. Its linear graph convolutional layer definition is as follows:
\begin{align}\label{eq:lgc}
\mathbf{E}(l+1)=\tilde{\mathbf{A}}\mathbf{E}(l),
\end{align} where $\mathbf{E}(0)\in \mathbb{R}^{(|\mathcal{U}| \times |\mathcal{V}|)\times D}$ is the learnable initial embedding matrix of users and items, $\mathbf{E}(l)$ denotes the embedding matrix at $l$-th layer, and $\tilde{\mathbf{A}}$ is the normalized user-item adjacency matrix. LightGCN learns the initial embedding and uses the layer combination. The model prediction is defined as the dot product of the user’s and item’s final representation $\bm{e}_u^T\bm{e}_i$. 

Other variants of LightGCN also achieved competitive performance~\cite{Xiang2020DGCF19,Mao21UltraGCN,Wu2021SGLED,mao2021simplex,fu2022revisiting,lee2021BUIR,hu2022mgdcf}. For example, SGL-ED~\cite{Wu2021SGLED} contrasts different node views that are generated by randomly masking the edge connections on the graph and incorporating the proposed self-supervised loss into LightGCN. DGCF~\cite{Xiang2020DGCF19} considers user-item relationships at the finer granularity of user intents and generated disentangled user and item representations. UltraGCN~\cite{Mao21UltraGCN} proposes a simplified CF that skips infinite layers of message passing for an efficient recommendation, which generalizes multiple standard linkage scores. SimpleX~\cite{mao2021simplex} improves the CF methods with the help of an appropriate negative sampling rate and proposed cosine contrastive loss. LinkProp~\cite{fu2022revisiting} proposes a new linkage score for link prediction on a bipartite graph. MGDCF~\cite{hu2022mgdcf} generalizes LightGCN and APPNP~\cite{Klicpera2019APPNP} with the Markov process for distance learning. GTN~\cite{fan2022GTN} captures the adaptive reliability of the interactions between users and items using the trend filter.

Recently, researchers argued that linear GCN-based models resemble heat equations, which describe the law of thermal diffusive processes, i.e., Newton’s Law of Cooling~\cite{choi2021ltocf, wang2021dgc}. LT-OCF~\cite{choi2021ltocf} redesigned LightGCN as a continuous diffusive process and outperforms LightGCN. LT-OCF also learns an optimal layer combination rather than relying on a pre-defined architecture. The heat equation is directly related to low pass filters and smoothness, which is one of the key operations in graph signal processing. GF-CF~\cite{Shen21GFCF} was proposed from the perspective of the smoothness of graph signals. It is the special case of existing CF methods: the low-rank matrix factorization corresponds to the ideal low-pass filter, and LightGCN with infinitely embedding dimensionality corresponds to a first-order linear filter. Therefore, GF-CF proposed a simple model combining a linear filter and an ideal low-pass filter as follows:
\begin{align}
     \hat{\bm{R}}= \bm{R}\big(\tilde{\bm{P}} + \beta \bm{V}^{-\frac{1}{2}} \bar{\bm{U}} \bar{\bm{U}}^{\mathsf{T}}\bm{V}^{\frac{1}{2}} \big),
\end{align}where $\hat{\bm{R}}$ is an inferred interaction matrix, and $\bar{\bm{U}}$ is the top-$k$ singular vectors of $\tilde{\bm{R}}$. GF-CF only needs matrix multiplication operations to calculate the recommendation scores thanks to its non-parametric architecture. Both LT-OCF and GF-CF use blurring processes with the heat equation and the low-pass filter, respectively, but no sharpening processes. In Table~\ref{tbl:cmp1}, we compare recent methods.

\begin{figure}
    \centering
    \includegraphics[width=1\columnwidth]{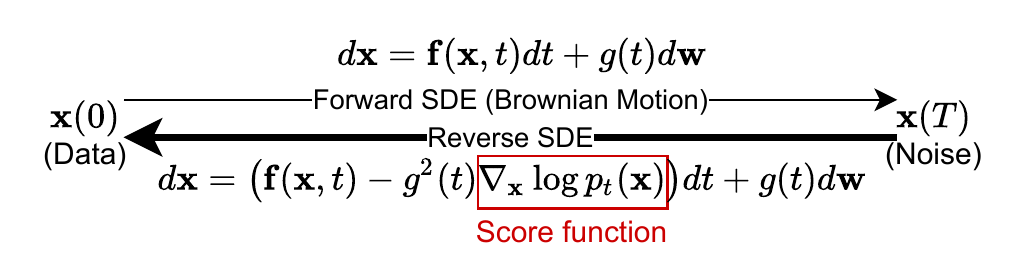}
    \caption{The overall workflow of SGMs, where the score function is approximated by a score network, i.e., $S_{\theta}(\mathbf{x},t) \approx \nabla_\mathbf{x}\log p_t(\mathbf{x})$. We note that it means the gradient of the log probability w.r.t. $\mathbf{x}$ at time $t$.}
    \label{fig:sde}
\end{figure}

\subsection{Score-based Generative Models (SGMs)}
Fig.~\ref{fig:sgm} (a) depicts the basic mechanism behind SGMs~\cite{song2021scorebased,song2021maximum,song2020improved}. The forward process is written as the following stochastic differential equation:
\begin{align}\label{eq:forward}
d\mathbf{x}=\mathbf{f}(\mathbf{x},t)dt + g(t)d\mathbf{w},
\end{align}where $\mathbf{f}(\mathbf{x},t) = f(t)\mathbf{x}$, and its reverse SDE (i.e., backward process) is defined as follows:
\begin{align}\label{eq:reverse}
d\mathbf{x}=\big(\mathbf{f}(\mathbf{x},t)-g^2(t)\nabla_\mathbf{x} \log p_t(\mathbf{x})\big)dt + g(t)d\mathbf{w},
\end{align}where this reverse SDE process is a generative process. 
Depending on the types of $f$ and $g$, various sub-types of SGMs are defined.

In order to solve the reverse SDE, we need to know the gradient of the log-probability of the forward SDE (i.e., $\nabla_\mathbf{x} \log p_t(\mathbf{x})$). We typically train a neural network, called \emph{score network}, to approximate it (cf. Fig.~\ref{fig:sde}). There exists a well-established theory for training the score network. After training the score network with the data collected during the forward process, one can generate fake data from noisy vectors using only the reverse SDE. We formally compare our BSPMs with SGMs as follows since they share similar processing philosophy:
\begin{enumerate}
    \item SGMs are for images. One image dataset includes many images and therefore, the entire process should be described in stochastic differential equations (SDEs).
    \item BSPMs deal with the user-item interaction matrix. One CF dataset includes only one such matrix and therefore, the entire process can be described by deterministic ordinary differential equations (ODEs).
    \item In both models, we expect that new information is discovered during the recovery process. For instance, the denoising process is a generative process in SGMs and in our case, user-specific items are recommended during the sharpening process.
    \item Except that a series of perturbation-recovery processes are used in both models, however, they differ at many detailed points. In BSPMs, most importantly, there does not exist anything to learn since we directly blur and sharpen the interaction matrix $\bm{R}$.
    \item In Table~\ref{tbl:cmp}, we summarize key differences.
\end{enumerate}
\begin{table}[t]
    \small
    \setlength{\tabcolsep}{2pt}
    \caption{The comparison between SGMs and our BSPMs\label{tbl:cmp}}
    \begin{tabular}{ccc}\toprule
         & SGM & Our proposed BSPM \\ \midrule
        Type & SDEs & ODEs \\ 
        Perturbation & Adding noises to images & Blurring interaction matrix \\
        Recovery & Denoising images & Sharpening blurred matrix \\ 
        Data & Many images  & One interaction matrix \\
        What to Learn & Score function & N/A \\         \bottomrule
    \end{tabular}
\end{table}


\subsection{Ordinary Differential Equations (ODEs)}\label{sec:solver}
The initial value problem (IVP) of ordinary differential equations can be written as follows:
\begin{align}
    \bm{x}(T) = \bm{x}(0) + \int_0^T f(\bm{x}(t)) dt,
\end{align}where $\bm{x}(0)$ is an initial value at time $t=0$, and $f:\mathbb{R}^{\dim(\bm{x})} \rightarrow \mathbb{R}^{\dim(\bm{x})}$ is an ODE function describing the time-derivative of $\bm{x}$, denoted by $\frac{d\bm{x}(t)}{dt}$\footnote{In the case of neural ordinary differential equations (NODEs), $f$ is approximated by a neural network, which means the time-derivative of $\bm{x}$ is learned from data [node]. In this paper, however, $f$ is not a neural network but a blurring/sharpening function.}. Therefore, integrating the time-derivative of $\bm{x}$ until $t=T$ returns a solution $\bm{x}(T)$ at time $t=T$.

$f$ is typically complicated in real-world applications and it is frequently impossible to find an analytical solution of $\bm{x}(T)$. We then typically use ODE solvers, such as the Euler method, the Runge-Kutta method, the Dormand--Prince (DOPRI) method, and so on~\cite{dormand1980dopri}. The Euler method \footnote{Note that Eq.~\eqref{eq:euler} is identical to a residual connection when $s=1$ and therefore, NODEs are a continuous generalization of residual networks.} is written as follows:
\begin{align}\label{eq:euler}
\bm{x}(t + s) = \bm{x}(t) + \tau \cdot f(\bm{x}(t)),
\end{align}where $\tau$ is a pre-configured step size.

Other ODE solvers use more complicated methods to update $\bm{x}(t + \tau)$ from $\bm{x}(t)$. For instance, the fourth-order Runge--Kutta (RK4) method uses the following method:
\begin{align}\label{eq:rk4}
\bm{x}(t + \tau) = \bm{x}(t) + \frac{s}{6}\Big(f_1 + 2f_2 + 2f_3 + f_4\Big),
\end{align}where $f_1 = f(\bm{x}(t))$, $f_2 = f(\bm{x}(t) + \frac{\tau}{2}f_1)$, $f_3 = f(\bm{x}(t) + \frac{\tau}{2}f_2)$, and $f_4 = f(\bm{x}(t)+\tau f_3)$.

In order to solve the above integral problem, therefore, we need to iterate one of the fixed-step ODE solvers $\lceil T/\tau \rceil$ times since each iteration updates $\bm{x}(t)$ to $\bm{x}(t+\tau)$. However, the DOPRI method is an adaptive solver, which dynamically adjusts the step-size $\tau$ depending on estimated potential errors. Therefore, the number of iterations is not deterministic for DOPRI. In general, DOPRI is considered one of the most advanced solvers. 
These solvers are already implemented on many deep learning platforms, such as PyTorch and TensorFlow. We test all those solvers for our experiments.

\section{Proposed Method}
We describe our BSPMs for CF, which consist of a blurring process and a sharpening process. Our method is greatly inspired by the recent successes of SGMs for deep generative tasks. In fact, there already exists a research trend to use generative models for CF due to the similarity in them~\cite{Wang2017IRGAN,Chae2018CFGAN,Wang2018GraphGAN,Chae2019RAGANBT,Wang2019AugCF,Sun2020LARA,Chen2021TagRec} --- revealing hidden interactions between users and items means that we generate new interactions.

\subsection{Overall Workflow}\label{sec:overall}
Our overall workflow is as simple as i) applying a continuous blurring process to the interaction matrix $\bm{R} $ to derive its blurred matrix $\bm{B}(T_b)$, and then ii) applying a continuous sharpening process to the blurred matrix to derive its sharpening matrix $\bm{S}(T_s)$. After these processes, we can recommend items as we will shortly describe.

We also make it clear that in our method, there does not exist anything to train. During the processes, neural networks are not used at all and we do not learn user/item embedding vectors. The blurring and sharpening functions are all hand-crafted functions (without any trainable parameters) in our method. Therefore, our process is surprisingly simple and the overall computation can be done quickly. However, our method outperforms all existing popular methods by non-trivial margins.

\paragraph{Meaning of Blurring} The blurring process is a core of CF. Many graph-based CF methods use graph convolutional filters that correspond to blurring processes~\cite{He20LightGCN,Shen21GFCF,balcilar2021analyzing}. In general, popular items are recommended to users after this process.

\paragraph{Meaning of Sharpening} The sharpening process is an inverse of the blurring and therefore, it retrieves user-specific items --- for general GCNs, similar sharpening processes are used to emphasize differences among node features~\cite{Bo2021fagcn,chien2021GPRGNN}. As we will show in our experiment section, it actually increases the overall recommendation accuracy while mostly decreasing the degree of recommended items. In other words, less popular items are also recommended to users in conjunction with popular items.

\subsection{Blurring Process}
Blurring processes can be written as follows and solved by the ODE solvers we reviewed in Sec.~\ref{sec:solver}:
\begin{align}\label{eq:blur}
    \bm{B}(T_b) = \bm{B}(0) + \int_0^{T_b} b(\bm{B}(t)) dt,
\end{align}where $b:\mathbb{R}^{\dim(\bm{B})} \rightarrow \mathbb{R}^{\dim(\bm{B})}$ is a blurring function which approximates $\frac{d\bm{B}(t)}{dt}$ and $\bm{B}(0)$ is an interaction matrix $\bm{R}$ in our setting. Therefore, $\bm{B}(1)$ means a blurred interaction matrix.

The exact blurring process depends on how we define the function $b$. In various domains, similar blurring functions have been defined for various purposes. We introduce some key definitions among them that are widely used in various domains.

We first articulate that all the aforementioned notations can be naturally extended after considering the temporal nature of our proposed blurring-sharpening process. For instance, $\bm{B}(t)$ means a blurred matrix of the original interaction matrix after $t$ following our proposed blurring process \textit{iff} $\bm{B}(0) = \bm{R}$.


\begin{figure}[t]
    \centering
    \subfigure[This model corresponds to GF-CF when we use i) the Euler method with $T_b=1$ and $\tau=1$ to solve Eq.~\eqref{eq:blur}, and ii) a heat capacity of $k=1$.]{\includegraphics[width=0.75\columnwidth]{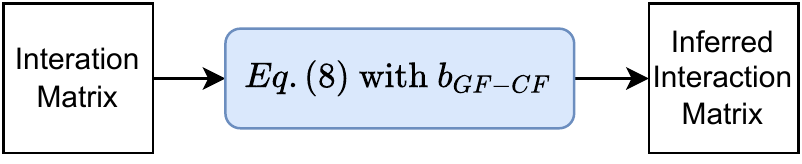}}
    \subfigure[The late-merge, denoted BSPM-LM, merges the ideal low-pass filter at the last moment.]{\includegraphics[width=\columnwidth]{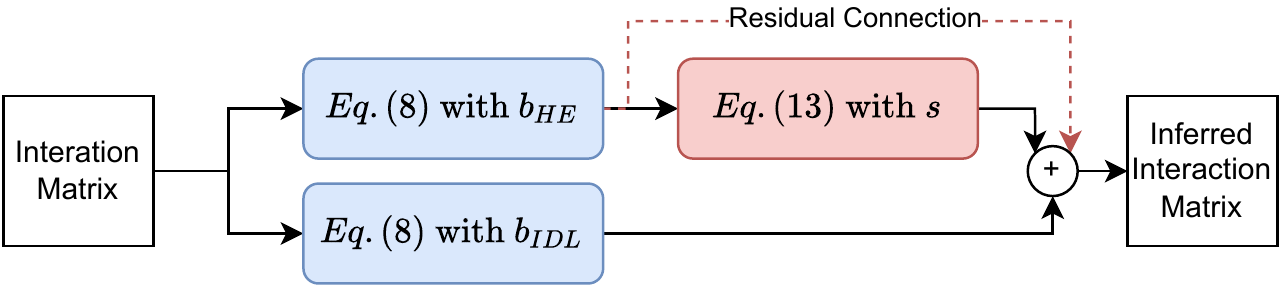}}
    \subfigure[The early-merge, denoted BSPM-EM, merges the ideal low-pass filter before the sharpening process begins.]{\includegraphics[width=\columnwidth]{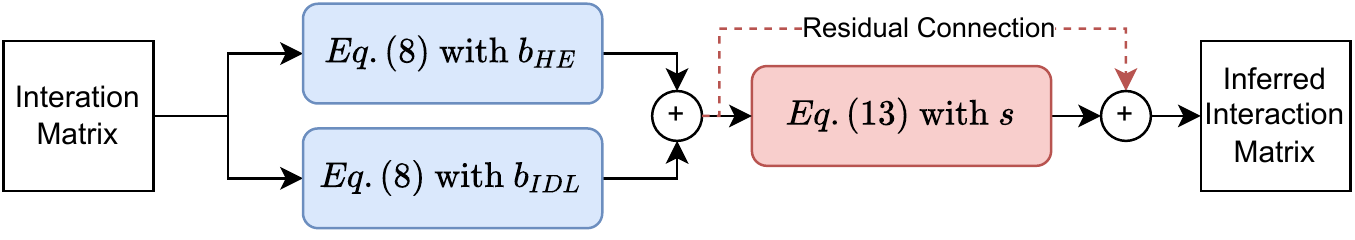}}
    \caption{Blue boxes mean blurring processes, and red boxes mean sharpening processes. The red dotted path means the residual connection, which is optional in our method. The residual connection enhances the recommendation accuracy in one dataset in our experiments.}
    \label{fig:bspm}
\end{figure}

\paragraph{Heat equation} The heat equation means the Newton's law of cooling, which describes the rate of heat loss in a body. This concept is frequently used in image processing for blurring images. We use the following definition of $b$:
\begin{align}
    b_{HE}(\bm{B}(t)) = k\bm{B}(t)\big(\tilde{\bm{P}} - \bm{I}\big),
\end{align}where $k \in \mathbb{R}$ is a coefficient called heat capacity. This is a hyperparameter in our framework. This definition of $b$ has a resemblance to the low-pass filter in the field of graph convolutions.


\paragraph{Ideal low-pass filter} In the field of graph convolutions, the following ideal low-pass filter is frequently used:
\begin{align}
    b_{IDL}(\bm{B}(t)) = \bm{B}(t)\big(\bm{V}^{-\frac{1}{2}} \bar{\bm{U}} \bar{\bm{U}}^{\mathsf{T}}\bm{V}^{\frac{1}{2}}-\bm{I}\big),
\end{align}where $\bar{\bm{U}}$ is the top-$k$ singular vectors of $\tilde{\bm{R}}$.

Our framework has a flexibility that one can also combine them, for instance, as follows:
\begin{align}\label{eq:gfcf}
    b_{GF-CF}(\bm{B}(t)) = k\bm{B}(t)\big(\tilde{\bm{P}} +\beta \bm{V}^{-\frac{1}{2}} \bar{\bm{U}} \bar{\bm{U}}^{\mathsf{T}} \bm{V}^{\frac{1}{2}} - \bm{I} \big),
\end{align}where $\beta$ is a coefficient to (de-)emphasize the ideal low-pass filter.

In particular, Eq.~\eqref{eq:blur} with $b_{GF-CF}$ reduces to GF-CF~\cite{Shen21GFCF} which can be written as follows and solved by the ODE solvers we reviewed in Sec.~\ref{sec:solver}:
\begin{align}\begin{split}\label{eq:gfcf2}
    \hat{\bm{R}} &= \bm{B}(0) + \int_0^{T_b} b_{GF-CF}(\bm{B}(t)) dt,\\
\end{split}\end{align}where $\bm{B}(0) = \bm{R}$, $k=1$, $T_b=1$, $\hat{\bm{R}}$ is an inferred interaction matrix, and we use the Euler method with $\tau = 1$. Therefore, one can consider that GF-CF is a CF method based only on the blurring process (cf. Fig.~\ref{fig:bspm} (a)). 

However, our work shows that it is sub-optimal to use only the blurring process. The following sharpening process is able to further enhance the recommendation accuracy. To our knowledge, we are the first proposing the blurring-sharpening process-based CF method.

\subsection{Sharpening Process}
Sharpening processes can also be written as follows and solved by the ODE solvers we reviewed in Sec.~\ref{sec:solver}:
\begin{align}\label{eq:sharp}
    \bm{S}(T_s) = \bm{S}(0) + \int_0^{T_s} s(\bm{S}(t)) dt,
\end{align}where $s:\mathbb{R}^{\dim(\bm{S})} \rightarrow \mathbb{R}^{\dim(\bm{S})}$ is a sharpening function which approximates $\frac{d\bm{S}(t)}{dt}$, and $\bm{S}(1)$ is a sharpened matrix from the input matrix $\bm{S}(0)$. The sharpening function $s$ can be defined as follows:
\begin{align}
    s(\bm{S}(t)) = -\bm{S}(t)\tilde{\bm{P}},
\end{align}where the negative sign is added to emphasize the difference from neighbors, i.e., sharpening.

\subsection{Blurring-Sharpening Process Model (BSPM)}
Using the blurring and sharpening processes, we can define a couple of variations of BSPM. In BSPM-LM in Fig.~\ref{fig:bspm} (b), we use the two blurring processes but apply the sharpening process only to the heat equation-based blurring outcome. We then merge the sharpened interaction matrix with the matrix perturbed by the ideal low-pass filter. This variant can be written as follows and solved by the ODE solvers we reviewed in Sec.~\ref{sec:solver}:
\begin{align}\begin{split}\label{eq:bspm1}
    \bm{B}_{HE}(T_b) &= \bm{B}(0) + \int_0^{T_b} b_{HE}(\bm{B}(t)) dt,\\
    \bm{B}_{IDL}(T_b) &= \bm{B}(0) + \int_0^{T_b} b_{IDL}(\bm{B}(t)) dt,\\
    \bm{S}(T_s) &= \bm{S}(0) + \int_0^{T_s} s(\bm{S}(t)) dt,\\
    \hat{\bm{R}} &= \begin{cases}
    \bm{S}(T_s) + \bm{B}_{IDL}(T_b) + \bm{B}_{HE}(T_b),\textrm{ if residual},\\
    \bm{S}(T_s) + \bm{B}_{IDL}(T_b),\textrm{ otherwise},
    \end{cases}
\end{split}\end{align}where $\bm{B}(0) = \bm{R}$, $\bm{S}(0) = \bm{B}_{HE}(T_b)$, and $\hat{\bm{R}}$ is an inferred interaction matrix. Adding $\bm{B}_{HE}(T_b)$ to $\hat{\bm{R}}$ is optional in our method and is called as \emph{residual connection}.

In BSPM-EM in Fig.~\ref{fig:bspm} (c), we merge the heat equation-based and the ideal low-pass filter-based blurring outcomes as early as before the sharpening process begins. We then apply the sharpening process. This can be written as follows and solved by the ODE solvers we reviewed in Sec.~\ref{sec:solver}:
\begin{align}\begin{split}\label{eq:bspm2}
    \bm{B}_{HE}(T_b) &= \bm{B}(0) + \int_0^{T_b} b_{HE}(\bm{B}(t)) dt,\\
    \bm{B}_{IDL}(T_b) &= \bm{B}(0) + \int_0^{T_b} b_{IDL}(\bm{B}(t)) dt,\\
    \bm{S}(T_s) &= \bm{S}(0) + \int_0^{T_s} s(\bm{S}(t)) dt,\\
    \hat{\bm{R}} &= \begin{cases}
    \bm{S}(T_s) + \bm{S}(0),\textrm{ if residual},\\
    \bm{S}(T_s),\textrm{ otherwise},
    \end{cases}
\end{split}\end{align}where $\bm{B}(0) = \bm{R}$, and $\bm{S}(0) = \bm{B}_{HE}(T_b) + \bm{B}_{IDL}(T_b)$. Adding $\bm{S}(0)$ to $\hat{\bm{R}}$ is a residual connection.


\subsection{Direct Inference without Training}
We note that our proposed BSPM does not include any training phase, which drastically reduces the total computation time. Since we do not learn any embedding vectors but directly process the interaction matrix $\bm{R}$, there is no training process. Solving Eq.~\eqref{eq:bspm1} or~\eqref{eq:bspm2} is enough to infer unknown user-item interactions, and there exist many ODE solvers which can solve Eqs.~\eqref{eq:bspm1} and~\eqref{eq:bspm2} efficiently. As a matter of fact, our method is one of the fastest CF methods. In addition, our method shows the best accuracy in almost all cases for our experiments.

\subsection{Comparison with Other Methods}
We already showed that GF-CF in Eq.~\eqref{eq:gfcf2} is a special case of BSPM. We will show that other popular CF algorithms are also special cases of our model: i) LightGCN is one of the most influential algorithms for linear graph-based CF. \citet{Shen21GFCF} already proved that LightGCNs with infinite-dimensional embeddings are theoretically the same as a one-step heat equation process, which is equivalent to our blurring process with $T_b=1$, the Euler method with a step size of 1. LightGCN also does not have any sharpening processes. ii) LT-OCF is a continuous generalization of LightGCN. Therefore, our ODE-based blurring process with the heat equation conceptually corresponds to the key idea of LT-OCF although it also has several other contributions. No sharpening processes are used in LT-OCF. iii) It is obvious that GF-CF is equivalent to the blurring process with $b_{GF-CF}$ and the Euler method of $T_b=1$ to solve it.

\section{Experiments}
In this section, we describe our experimental environments and results. The following software and hardware environments were used for all experiments: \textsc{Ubuntu} 18.04 LTS, \textsc{Python} 3.6.6, \textsc{PyTorch} 1.9.0, \textsc{Numpy} 1.18, \textsc{Scipy} 1.5, \textsc{sparsesvd} 0.2.2, \textsc{torchdiffeq} 0.2.2, \textsc{CUDA} 11.4, \textsc{NVIDIA} Driver 470.42, i9 CPU, and \textsc{RTX A6000}.
\subsection{Experimental Environments}

\subsubsection{Datasets and Baselines}

\begin{figure}[t]
    \centering
\includegraphics[width=0.65\columnwidth]{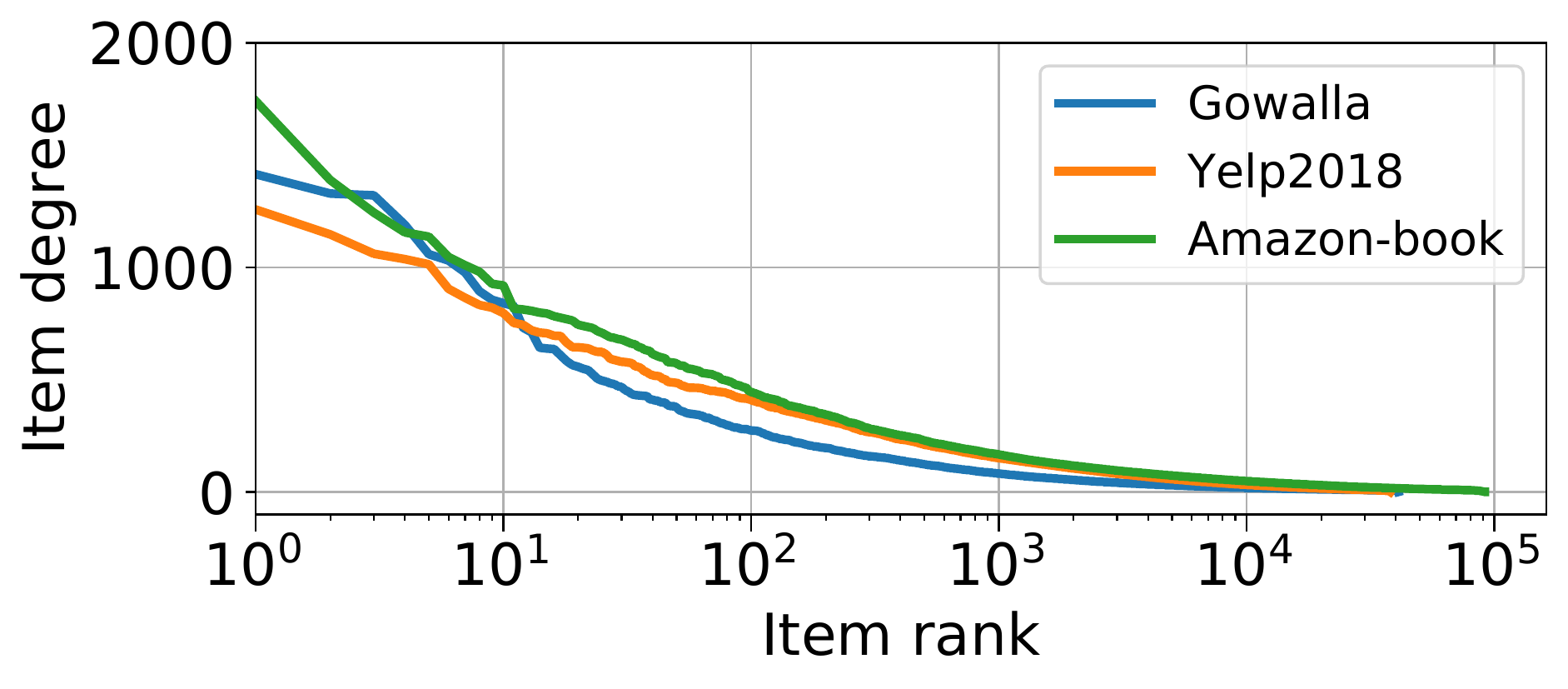}
    \caption{The long-tail characteristic in all datasets.}
    \label{fig:longtail}
\end{figure}

In our experiments, we use the three benchmark datasets that are the most frequently used in the literature: Gowalla, Yelp2018, and Amazon-book~\cite{Wang19NGCF,chen20LRGCCF,He20LightGCN}.  We summarize the dataset statistics in Table~\ref{tbl:data}. Fig.~\ref{fig:longtail} shows the long tail characteristic of the datasets. We compare our proposed BSPM with the following baseline models of different groups:

\begin{enumerate}
    \item In the first group of baselines, we consider popular MF-based methods and its variants: MF-BPR~\cite{rendle2009BPR}, Neu-MF~\cite{He2017NCF}, HOP-Rec~\cite{yang2018hop},  GRMF~\cite{Rao2015GRMF}, ENMF~\cite{chen2020ENMF}, and MF-CCL~\cite{mao2021simplex}.
    \item The second group includes autoencoder-based methods for CF: Mult-VAE~\cite{Liang2018VAECF}, Macrid-VAE~\cite{ma2019macridvae}, and EASE$^R$~\cite{steck2019EASE}.  
    \item The third group includes popular network embedding methods: DeepWalk~\cite{perozzi2014deepwalk}, LINE~\cite{tang2015line}, Node2Vec~\cite{grover2016node2vec}, and Item2Vec~\cite{Oren2016Item2Vec}.
    \item These three models are based on various deep learning paradigms: YoutubeNet~\cite{Covington2016YoutubeNet} is an MLP-based method, CML~\cite{hsieh2017cml} is a metric learning-based method, and CMN~\cite{Ebesu2018CMN} is a memory network-based model.
    \item The fifth group includes general GCN methods: GAT~\cite{velickovic2018GAT}, JKNet~\cite{xu2018jknet}, APPNP~\cite{Klicpera2019APPNP}, DisenGCN~\cite{ma2019DisenGCN}, and DropEdge~\cite{rong2020dropedge}.
    \item The last group includes GCN-based CF methods: GC-MC~\cite{Rianne2017GCMC}, NGCF~\cite{Wang19NGCF}, LR-GCCF~\cite{chen20LRGCCF},  LightGCN~\cite{He20LightGCN}, NIA-GCN~\cite{sun2020NIA-GCN}, DeosGCF~\cite{liu2020deoscillated}, IMP-GCN~\cite{liu2021IMP-GCN}, SGL-ED~\cite{Wu2021SGLED}, HMLET~\cite{kong2022hmlet}, DGCF~\cite{Xiang2020DGCF19}, IA-GCN~\cite{zhang2022iagcn}, BUIR\textsubscript{NB}~\cite{lee2021BUIR}, UltraGCN~\cite{Mao21UltraGCN}, SimpleX~\cite{mao2021simplex}, LT-OCF~\cite{choi2021ltocf}, GF-CF~\cite{Shen21GFCF}, LinkProp~\cite{fu2022revisiting}, MGDCF~\cite{hu2022mgdcf}, and GTN~\cite{fan2022GTN}.
\end{enumerate}

\begin{table}[t]
    \small
    \centering
    \caption{Statistics of datasets}\label{tbl:data}
    \begin{tabular}{ccccc}\toprule
        Dataset     & \#Users & \#Items & \#Interactions & Density \\ \midrule
        Gowalla     & 29,858  & 40,981  & 1,027,370      & 0.084\% \\
        Yelp2018    & 31,668  & 38,048  & 1,561,406      & 0.130\% \\
        Amazon-book & 52,643  & 91,599  & 2,984,108      & 0.062\% \\
        \bottomrule
    \end{tabular}
\end{table}

\subsubsection{Evaluation Metrics and Hyperparameters}
We adopt the two widely used ranking metrics: Recall@20 and NDCG@20~\cite{Jarvelin2002NDCG}. All items that do not have any interactions with a user are recommendation candidates for the user. To keep the comparison fair with previous studies, we use the same datasets and the same train/test splits.

For other baselines, we use the recommended hyperparameters and for our method, we test the following hyperparameters:
\begin{itemize}
    \item For solving the integral problems of the blurring/sharpening processes, we consider the following ODE solvers: the Euler method, RK4, and DOPRI. However, we found that RK4 and DOPRI produce almost the same results in our preliminary experiments so we test only the Euler method and RK4.
    \item For the blurring process, the number of steps $\frac{T_b}{\tau}$ for solvers is in \{1,2,3,4\}, and the terminal time $T_b$ is set to 1 to 5. 
    \item For the sharpening process, the number of steps $\frac{T_s}{\tau}$ is in \{1,2,3,4\}, and the terminal time $T_s$ is set to 1 to 5.
    \item The size of $\beta$ is in $\{0.0,0.1,\cdots,1.0\}$.
    \item The heat capacity $k$ is in $\{0.1,\cdots,1.0\}$.
\end{itemize}

Among the test configurations, the best configuration set in each data is as follows: In Gowalla, $\frac{T_b}{\tau} =1$, $T_b=1$, $\frac{T_s}{\tau}=1$, $T_s=2.5$,  $k=1.0$, and $\beta=0.2$. In Yelp2018, $\frac{T_b}{\tau} =1$, $T_b=1$, $\frac{T_s}{\tau}=1$, $T_s=1.2$, $k=1.0$, and $\beta=0.3$. In Amazon-book, $\frac{T_b}{\tau}=1$, $T_b=1$, $\frac{T_s}{\tau}=2$, $T_s=2.2$, $k=1.0$, and $\beta=0$. In general, it is the best to use the Euler method for the blurring process and RK4 for the sharpening process. However, for Yelp2018, it is the best to use the Euler method for sharpening.

\begin{table}[ht!]
    \small
    \setlength{\tabcolsep}{2pt}
    \centering
    \caption{Overall performance comparison. \textit{Relative improvement} stands for the improvement of BSPM against the second-best baseline.}
    \begin{tabular}{c cc cc cc}\toprule
        \multirow{2}{*}{Model}  & \multicolumn{2}{c}{Gowalla} & \multicolumn{2}{c}{Yelp2018} & \multicolumn{2}{c}{Amazon-book} \\ \cmidrule(lr){2-3}\cmidrule(lr){4-5}\cmidrule(lr){6-7}
                                & Recall & NDCG             & Recall & NDCG    & Recall & NDCG \\ \midrule
        MF-BPR                  & 0.1291 & 0.1109           & 0.0433 & 0.0354  & 0.0250 & 0.0196 \\
        GRMF                    & 0.1477 & 0.1205           & 0.0571 & 0.0462  & 0.0354 & 0.0270 \\
        GRMF-Norm               & 0.1557 & 0.1261           & 0.0561 & 0.0454  & 0.0352 & 0.0269 \\
        NeuMF                   & 0.1399 & 0.1212           & 0.0451 & 0.0363  & 0.0258 & 0.0200 \\
        HOP-Rec                 & 0.1399 & 0.1214           & 0.0517 & 0.0428  & 0.0309 & 0.0232 \\
        ENMF                    & 0.1523 & 0.1315           & 0.0624 & 0.0515  & 0.0359 & 0.0281 \\
        MF-CCL           & 0.1837 & 0.1493           & 0.0698 & 0.0572  & 0.0559 & 0.0447 \\\midrule
        Mult-VAE                & 0.1641 & 0.1335           & 0.0584 & 0.0450  & 0.0407 & 0.0315 \\
        Macrid-VAE              & 0.1618 & 0.1202           & 0.0612 & 0.0495  & 0.0383 & 0.0295 \\\
        EASE\textsuperscript{R}              & 0.1765	& 0.1467           & 0.0657	& 0.0552  & 0.0710 & 0.0567 \\\midrule
        YouTubeNet              & 0.1754 & 0.1473           & 0.0686 & 0.0567  & 0.0502 & 0.0388 \\
        CMN                     & 0.1405 & 0.1221           & 0.0475 & 0.0369  & 0.0267 & 0.0218 \\
        CML                     & 0.1670 & 0.1292           & 0.0622 & 0.0536  & 0.0522 & 0.0428 \\\midrule
        DeepWalk                & 0.1034 & 0.0740           & 0.0476 & 0.0378  & 0.0346 & 0.0264 \\
        LINE                    & 0.1335 & 0.1056           & 0.0549 & 0.0446  & 0.0410 & 0.0318 \\
        Node2Vec                & 0.1019 & 0.0709           & 0.0452 & 0.0350  & 0.0402 & 0.0309 \\
        Item2Vec                & 0.1325 & 0.1057           & 0.0503 & 0.0411  & 0.0326 & 0.0251 \\\midrule
        GAT                     & 0.1401 & 0.1236           & 0.0543 & 0.0431  & 0.0326 & 0.0235 \\
        JKNet                   & 0.1622 & 0.1391           & 0.0608 & 0.0502  & 0.0268 & 0.0343 \\
        DropEdge                & 0.1627 & 0.1394           & 0.0614 & 0.0506  & 0.0342 & 0.0270 \\
        APPNP                   & 0.1708 & 0.1462           & 0.0635 & 0.0521  & 0.0384 & 0.0299 \\
        DisenGCN                & 0.1356 & 0.1174           & 0.0558 & 0.0454  & 0.0329 & 0.0254 \\\midrule
        GC-MC                   & 0.1395 & 0.1204           & 0.0462 & 0.0379  & 0.0288 & 0.0224 \\
        PinSage                 & 0.1380 & 0.1196           & 0.0471 & 0.0393  & 0.0282 & 0.0219 \\
        NGCF                    & 0.1570 & 0.1327           & 0.0579 & 0.0477  & 0.0344 & 0.0263 \\
        NIA-GCN                 & 0.1359 & 0.1106           & 0.0599 & 0.0491  & 0.0369 & 0.0287 \\
        LR-GCCF                 & 0.1701 & 0.1452           & 0.0604 & 0.0498  & 0.0375 & 0.0296 \\
        LightGCN                & 0.1830 & 0.1554           & 0.0649 & 0.0530  & 0.0411 & 0.0315 \\
        SGL-ED                  & 0.1835 & 0.1539           & 0.0675 & 0.0555  & 0.0478	& 0.0379 \\
        DeosGCF                 & 0.1784 & 0.1477           & 0.0626 & 0.0504  & 0.0410 & 0.0316 \\
        IMP-GCN                 & 0.1845 & 0.1567           & 0.0653 & 0.0531  & 0.0460 & 0.0357 \\
        {BUIR\textsubscript{NB}}             & 0.1575 & 0.1301           & 0.0647 & 0.0526  & 0.0439 & 0.0346 \\
        DGCF                    & 0.1842 & 0.1561           & 0.0654 & 0.0534  & 0.0422 & 0.0324 \\
        {IA-GCN}                  & 0.1839 & 0.1562           & 0.0659 & 0.0537  & 0.0472 & 0.0373 \\
        UltraGCN                & 0.1862 & 0.1580 & 0.0683 & 0.0561  & 0.0681 & 0.0556 \\
        {SimpleX}                 & 0.1872 & 0.1557           & 0.0701 & 0.0575  & 0.0583 & 0.0468 \\
        LT-OCF                  & 0.1875 & 0.1574           & 0.0671 & 0.0549  & 0.0442 & 0.0341 \\
        GF-CF                   & 0.1849 & 0.1518           & 0.0697 & 0.0571  & 0.0710 & 0.0584 \\
        HMLET            & 0.1874 & \underline{0.1589} & 0.0675 & 0.0557  & 0.0482 & 0.0371 \\
        LinkProp                & 0.1814 & 0.1477           & 0.0676 & 0.0559  & 0.0684 & 0.0559 \\
        LinkProp-Multi          & \underline{0.1908} & 0.1573
                                                            & 0.0690 & 0.0571  & \underline{0.0721} & 0.0588 \\
        MGDCF        & 0.1864 & \underline{0.1589}
                                                            & 0.0696 & 0.0572  & 0.0490 & 0.0378 \\
        GTN              & 0.1870 & 0.1588           & 0.0679 & 0.0554  & 0.0450 & 0.0346 \\
        \midrule
        Only Blurring (HE)      & 0.1682 & 0.1331           & 0.0684 & 0.0565  & 0.0710 & 0.0584 \\
        Only Blurring (IDL)     & 0.1776 & 0.1489           & 0.0668 & 0.0549  & 0.0395 & 0.0316 \\
        Only Blurring (GF-CF)   & 0.1854 & 0.1518           & 0.0701 & 0.0575 & 0.0710  & 0.0584\\
        \midrule
        \textbf{BSPM-LM}        & 0.1901 & 0.1570          & \underline{0.0713} & \underline{0.0584}  & \textbf{0.0733} & \textbf{0.0610}\\
        \textbf{BSPM-EM}        & \textbf{0.1920} & \textbf{0.1597}
                                                            & \textbf{0.0720} & \textbf{0.0593}  & \textbf{0.0733} & \underline{0.0609}\\
        \midrule
        \textit{Relative Improvement}    
                                & 0.63\% & 0.50\% & 2.71\% & 3.13\% & 1.66\% & 3.74\% \\
        \bottomrule
    \end{tabular}
    \label{tbl:main_exp}
\end{table}

\subsection{Experimental Results}
In Table~\ref{tbl:main_exp}, we summarize the overall accuracy in terms of Recall@20 and NDCG@20. Our specific choices of baselines cover almost all representative CF methods, and the three datasets are widely used in the literature. As reported, our method clearly marks the best accuracy in all cases. In particular, BSPM-LM is the best method and performs better than the LinkProp-Multi by 3.74\% on NDCG@20 for Amazon-books. BSPM-EM is the best method for Yelp2018 and Gowalla. In many cases, BSPM-LM and BSPM-EM mark the best and the second-best methods, respectively, and their differences are not significant. 

Among the tested baselines, LinkProp-Multi, SimpleX, and GF-CF work well in some cases. However, only LinkProp-Multi is comparable to our method for Gowalla and Amazon-book --- however, the accuracy gap between our method and  LinkProp-Multi is still non-trivial. For Yelp2018, SimpleX, GF-CF, and MGDCF show good scores. For Amazon-book, GF-CF, LinkProp-Multi, and EASE\textsuperscript{R} show high performance. However, no existing methods are comparable to our proposed method in all datasets. Therefore, we consider that our proposed concept of BSPM opens a new era of CF.

LightGCN is worse than recent methods such as LT-OCF and GF-CF, but its value is that it is the first method showing that simple linear convolutions work better than non-linear ones. Being inspired by it, many methods have been proposed, including ours. Our blurring and sharpening processes are all linear operations. One can adopt non-linear sharpening processes, but in general, linear operations show reliable recommendations.

\subsection{Ablation and Sensitivity Studies}
We report some selected key sensitivity and ablation study results. It is the case that our model is not significantly sensitive to a hyperparameter if not reported in this subsection. In general, our model is sensitive to the hyperparameters of the sharpening process and we focus on them.

\subsubsection{Sensitivity on $T_s$.} 
\begin{figure}[t]
    \centering
    \subfigure[Gowalla]{\includegraphics[width=0.325\columnwidth]{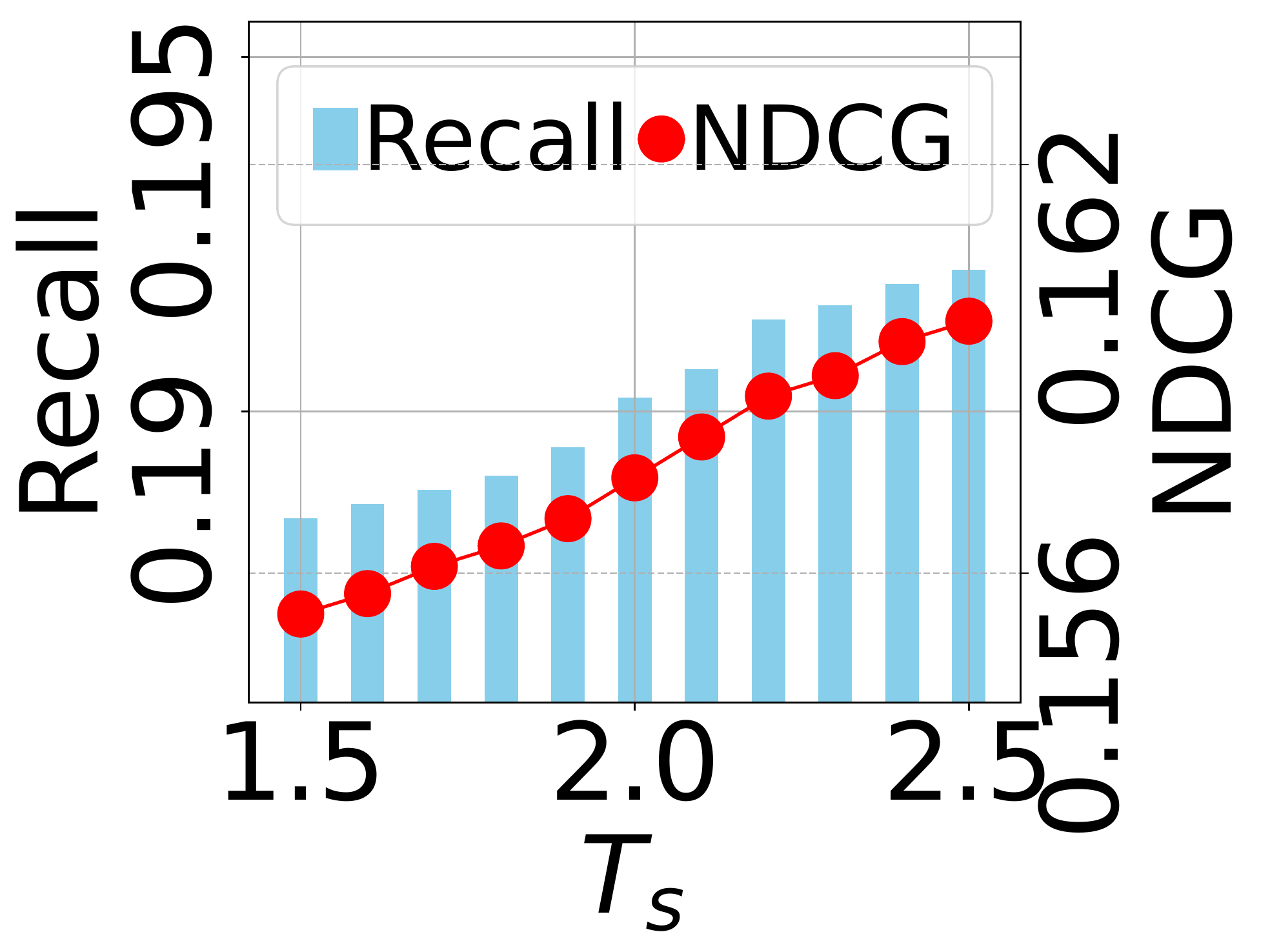}}
    \subfigure[Yelp2018]{\includegraphics[width=0.325\columnwidth]{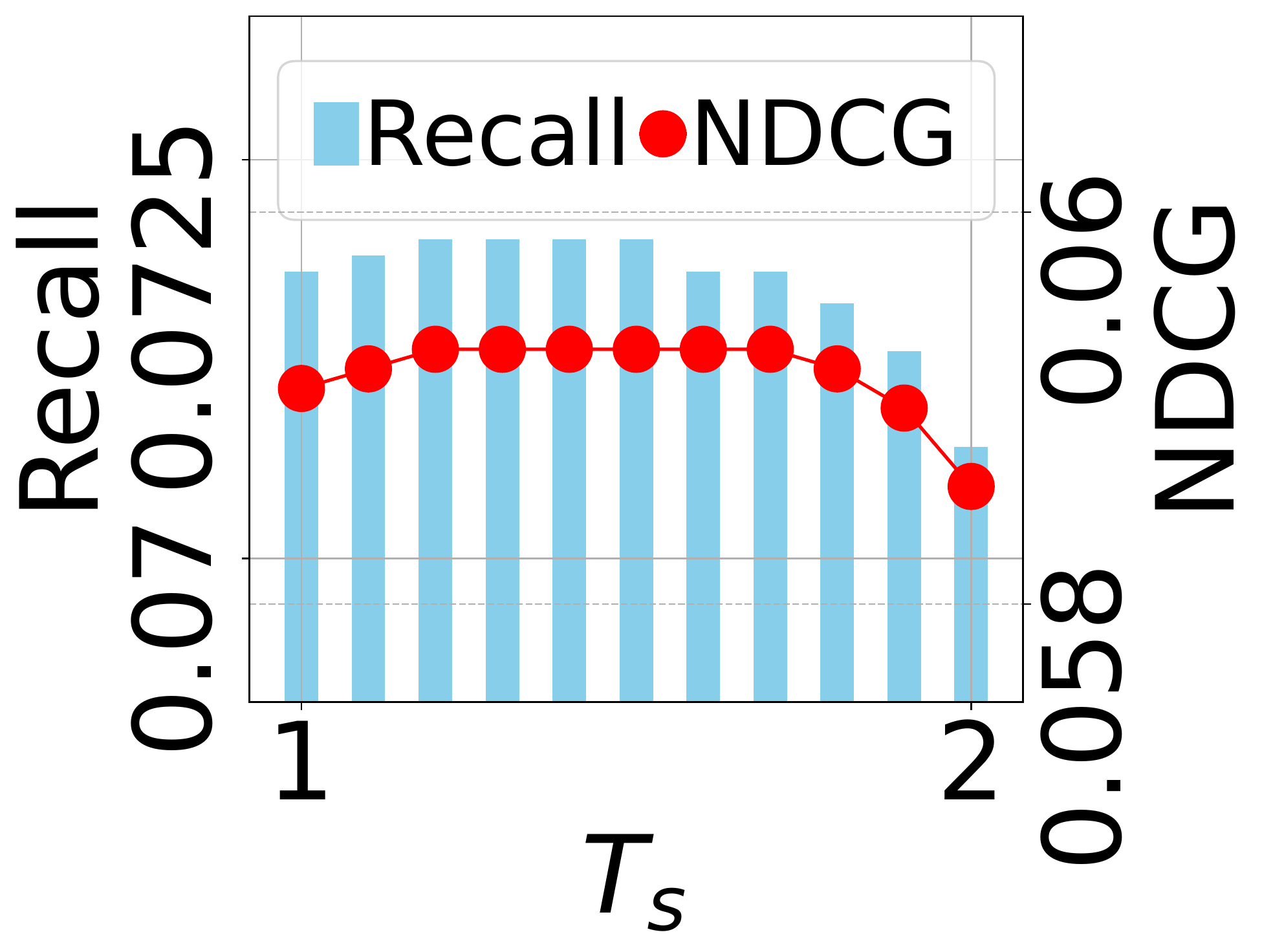}}
    \subfigure[Amazon-book]{\includegraphics[width=0.325\columnwidth]{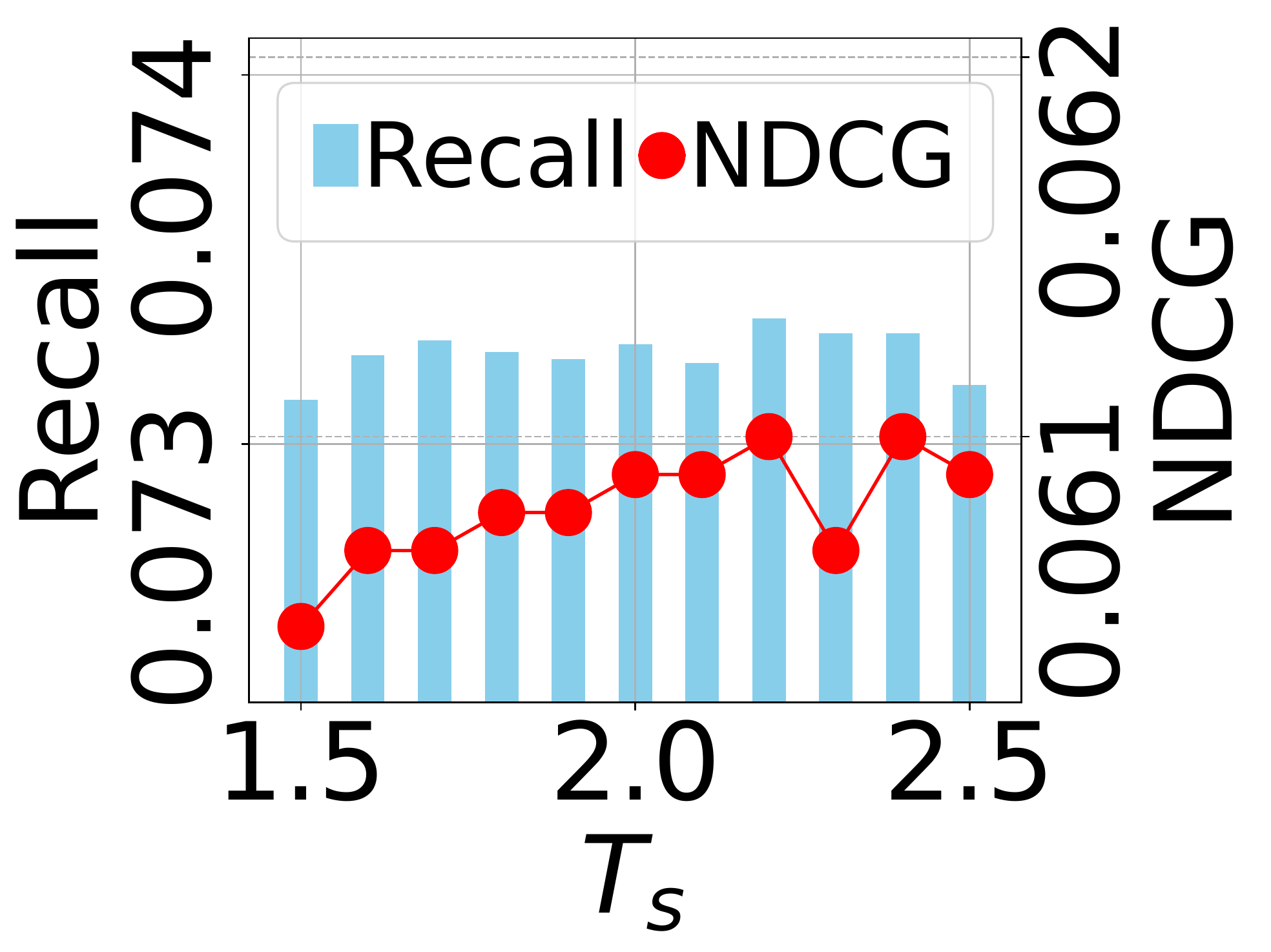}}
    \caption{Performance by varying $T_s$}
    \label{fig:terminal_time}
\end{figure}

By varying the terminal integral time $T_s$ of the sharpening process, we investigate how the model accuracy changes in Fig.~\ref{fig:terminal_time}. For Gowalla and Amazon-book, $T_s$ around 2.4 produces the best outcomes. After a certain point, however, the model accuracy drastically decreases in all datasets. It is obvious that applying the sharpening too much (i.e., $T_s$ is too large) is not helpful in the perspective of CF since the sharpening process emphasizes user-specific information (rather than collaborative information). In general, the blurring process can be considered as a collaborative step, where a user's interactions with items are mixed with its neighbors.

\subsubsection{Sensitivity on $\frac{T_s}{\tau}$.} 
\begin{figure}[t]
    \centering
    \subfigure[Gowalla]{\includegraphics[width=0.325\columnwidth]{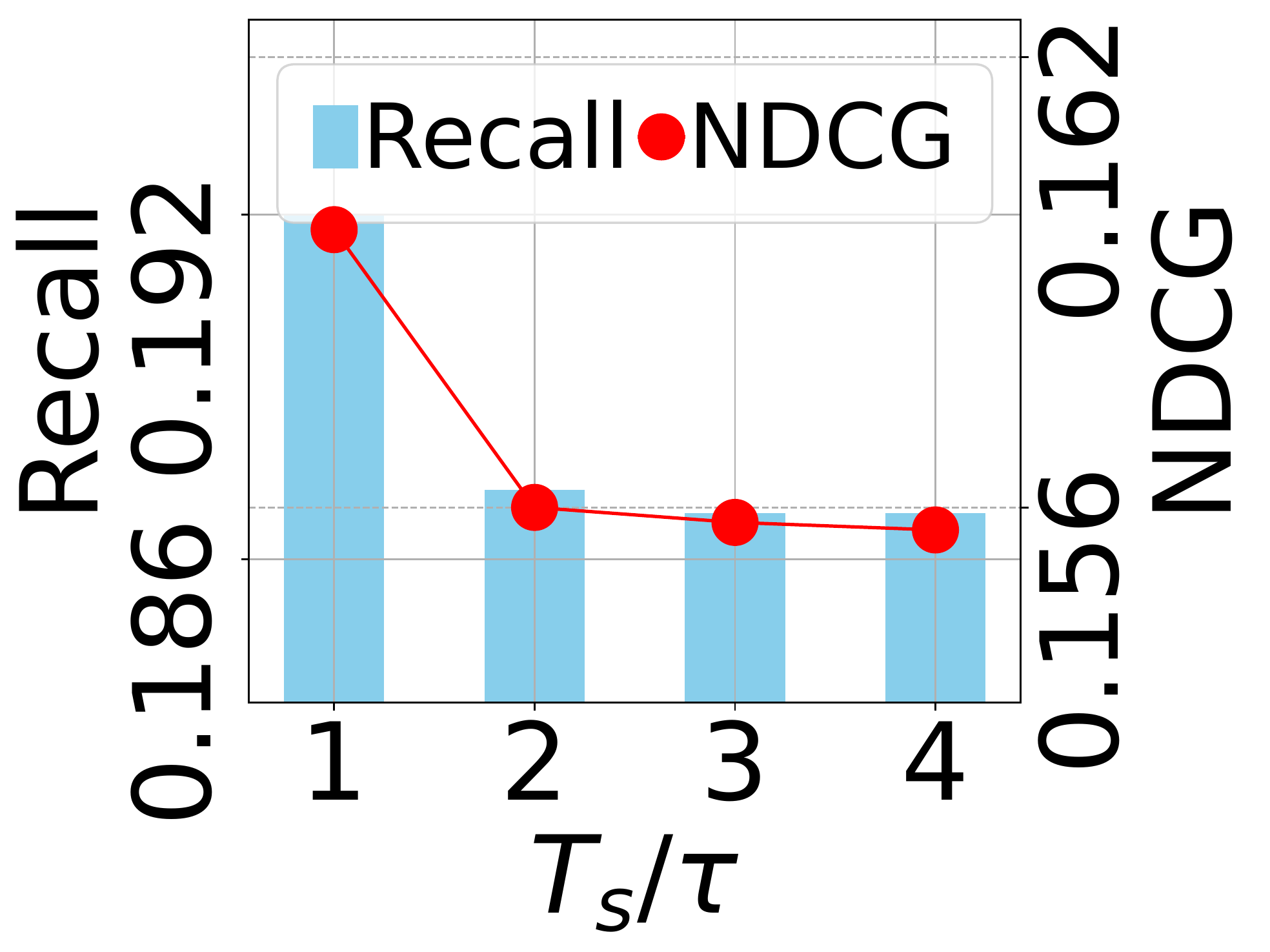}}
    \subfigure[Yelp2018]{\includegraphics[width=0.325\columnwidth]{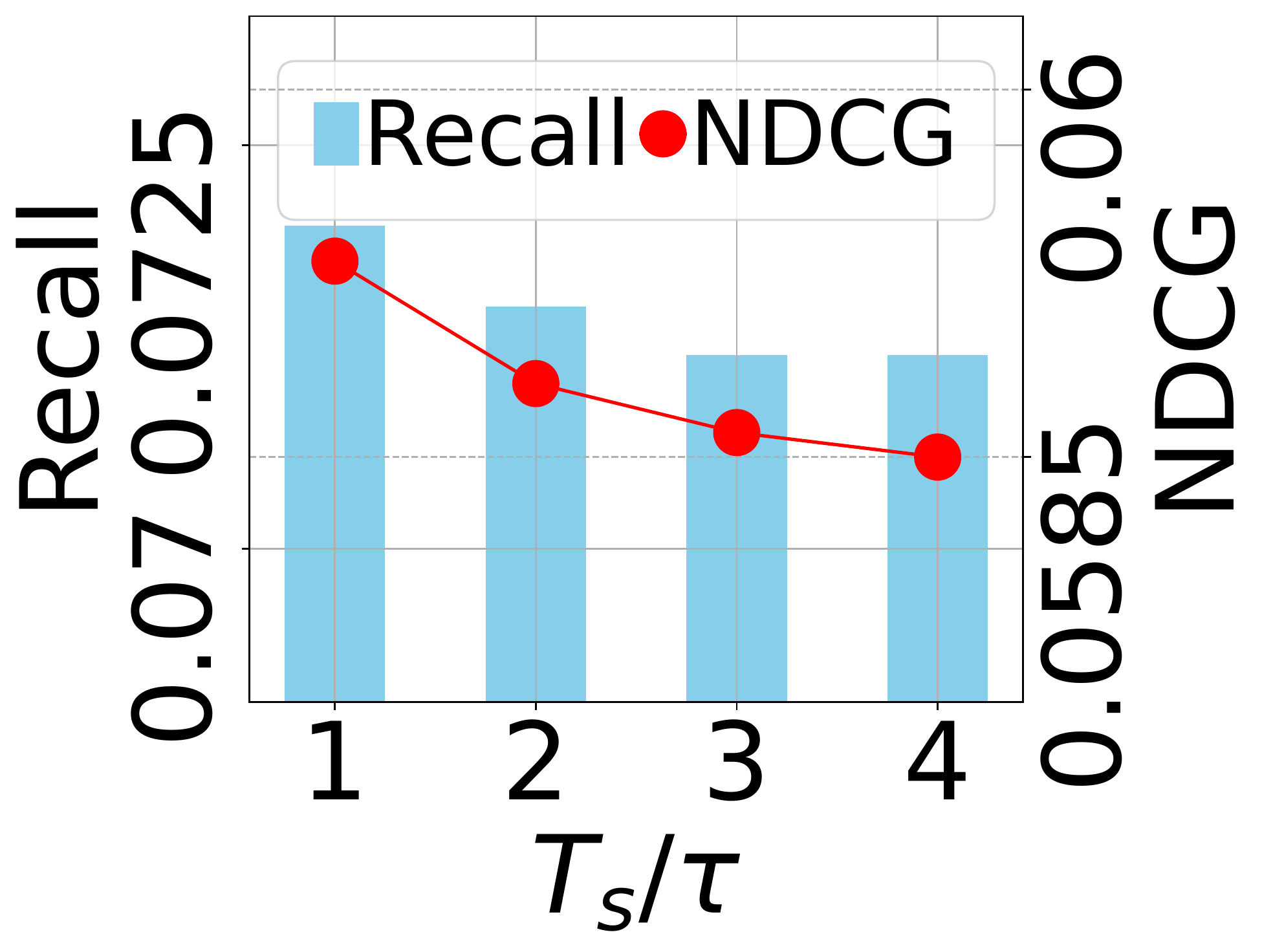}}
    \subfigure[Amazon-book]{\includegraphics[width=0.325\columnwidth]{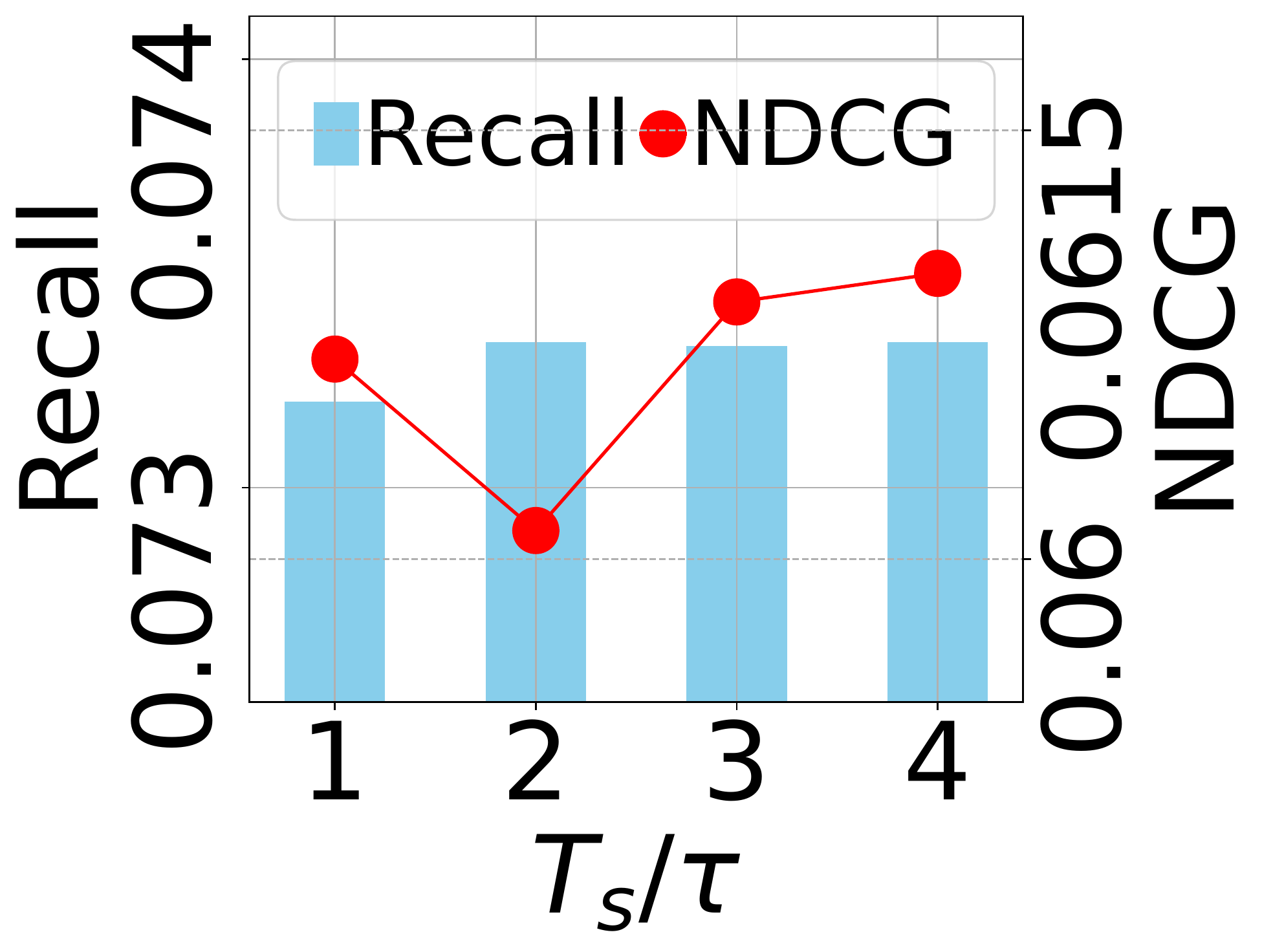}}
    \caption{Performance by varying $\frac{T_s}{\tau}$}
    \label{fig:steps}
\end{figure}

By varying the number of steps $\frac{T_s}{\tau}$ for ODE solvers, we test our model. Fig~\ref{fig:steps} shows that we do not need to use many steps in solving the integral problem of the sharpening process, i.e., Eq.~\eqref{eq:sharp}, which makes the overall runtime short. In most cases, it shows the best outcomes when $\frac{T_s}{\tau}$ is set to 1 or 2, i.e., $\tau$ is set to $T_s$ or $\frac{T_s}{2}$.


\begin{figure}[t]
    \centering
    \subfigure[Recall in Gowalla]{\includegraphics[width=0.49\columnwidth]{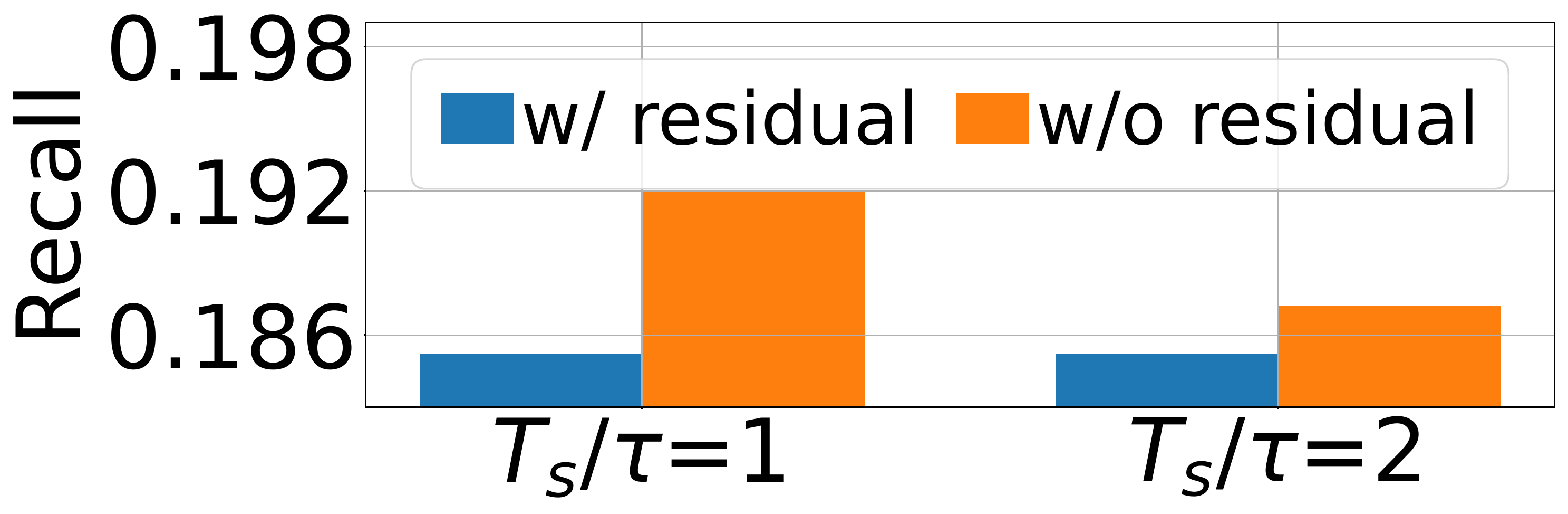}}\hfill
    \subfigure[NDCG in Gowalla]{\includegraphics[width=0.49\columnwidth]{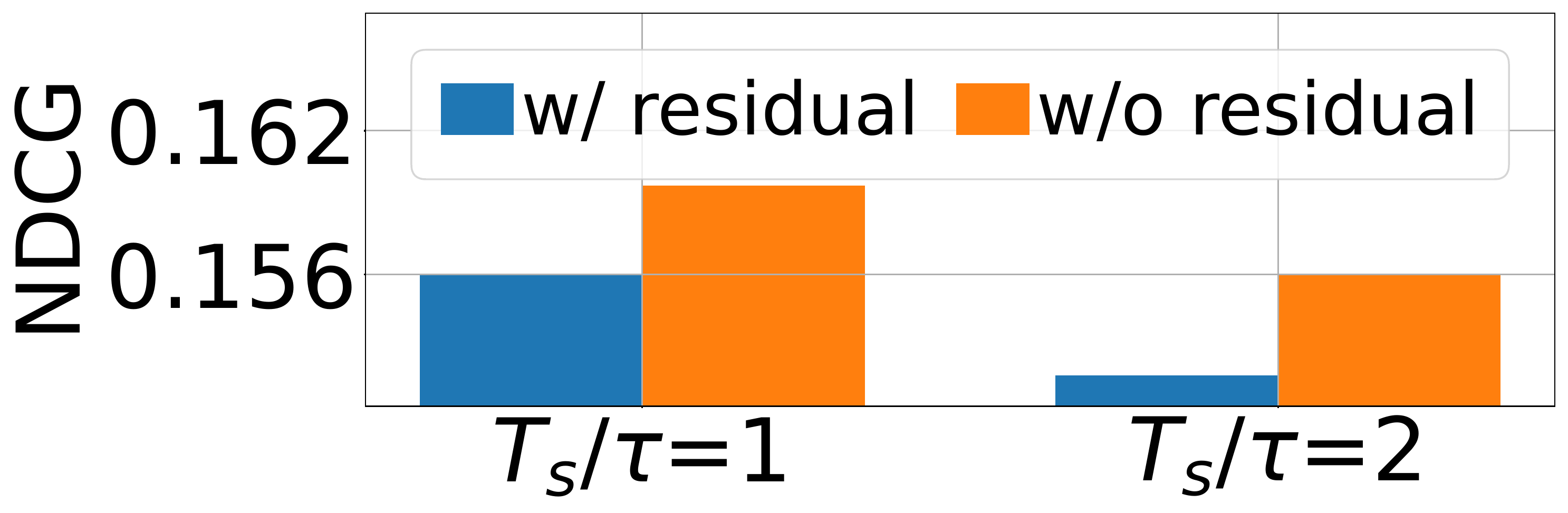}}\\
    \subfigure[Recall in Yelp2018]{\includegraphics[width=0.49\columnwidth]{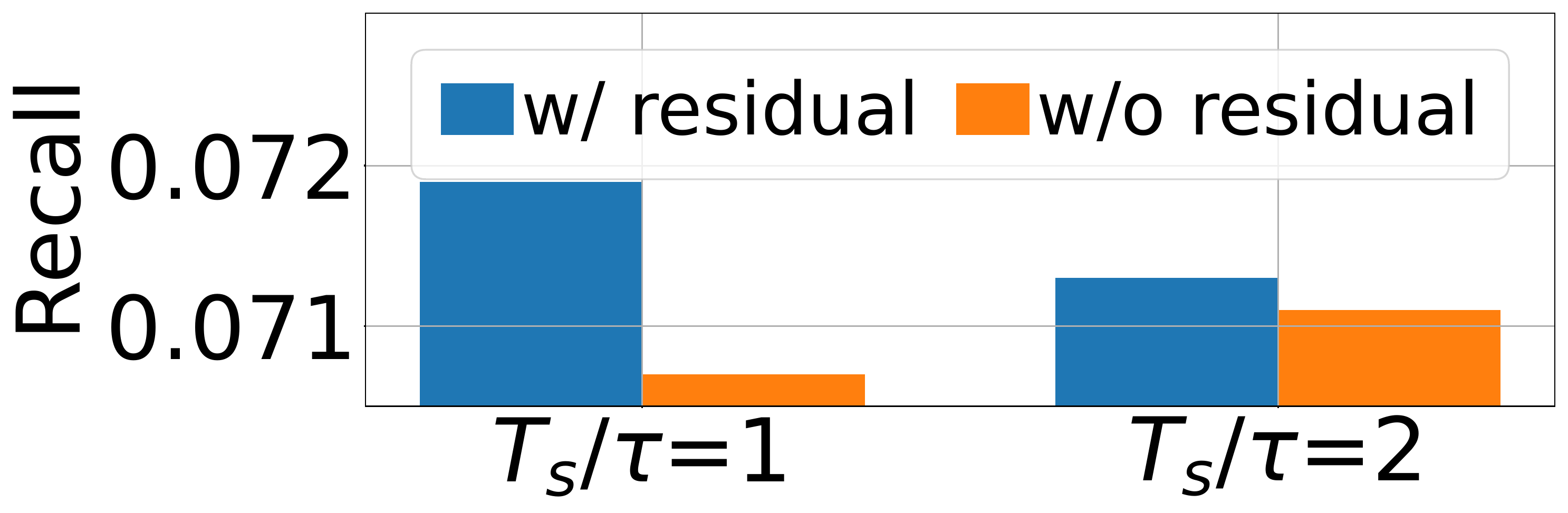}}\hfill
    \subfigure[NDCG in Yelp2018]{\includegraphics[width=0.49\columnwidth]{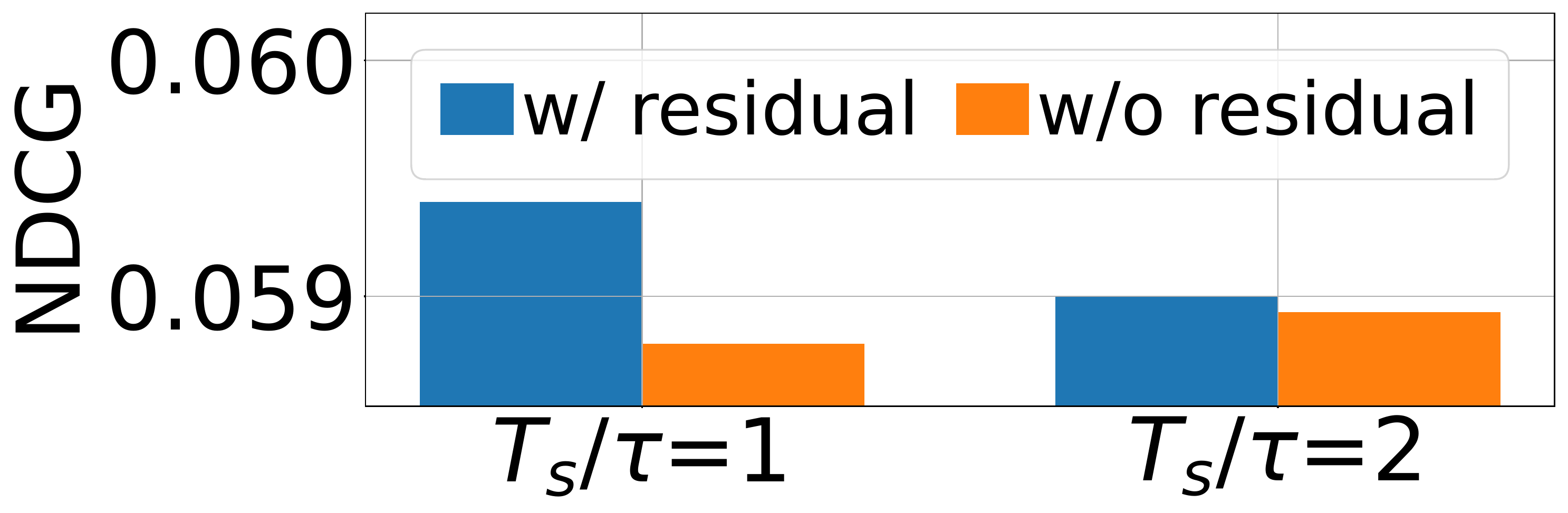}}\\
    \subfigure[Recall in Amazon-book]{\includegraphics[width=0.49\columnwidth]{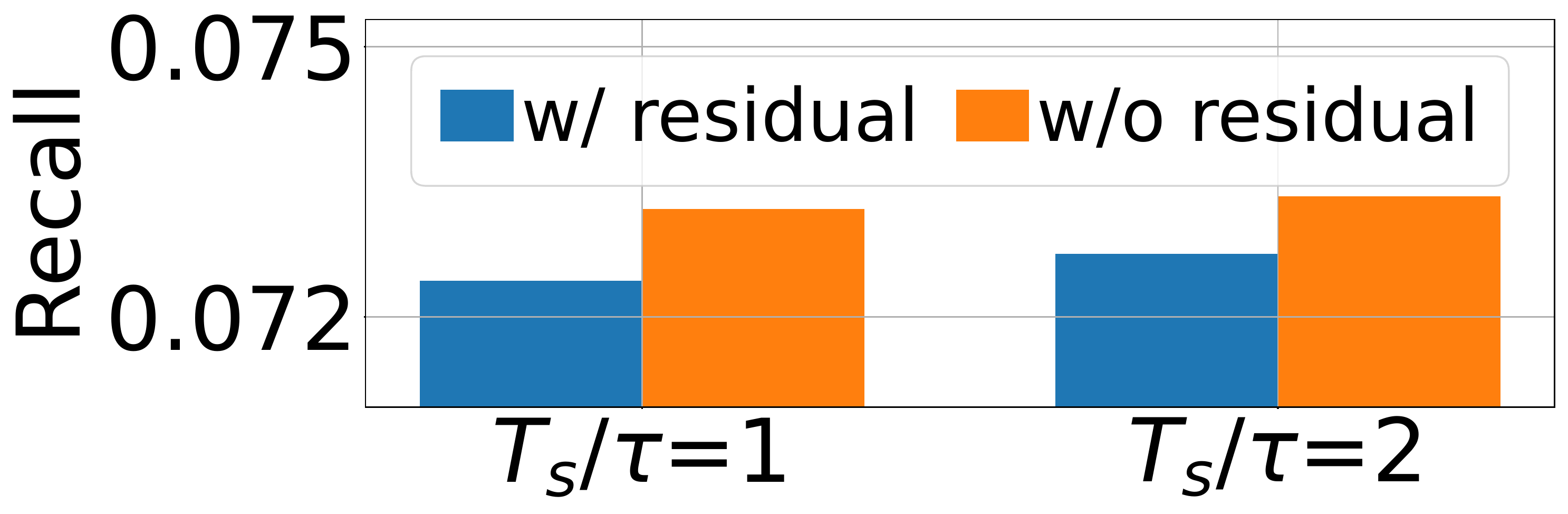}}\hfill
    \subfigure[NDCG in Amazon-book]{\includegraphics[width=0.49\columnwidth]{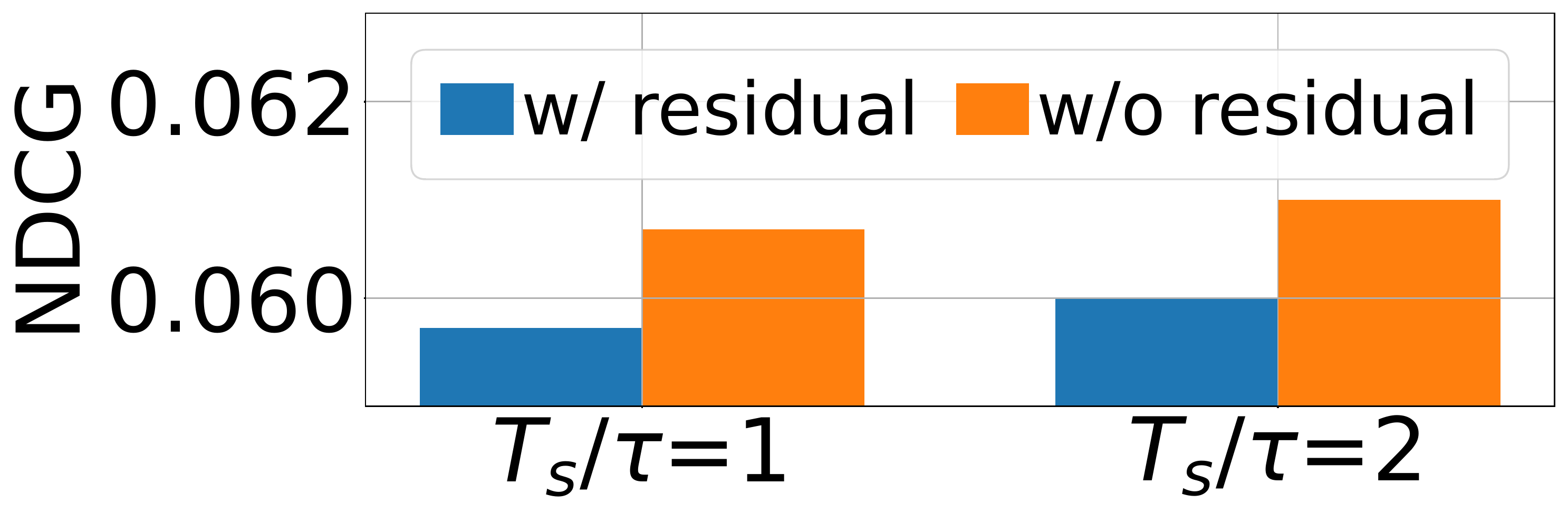}}
    \caption{Performance by the residual connection}
    \label{fig:combi}
\end{figure}

\begin{figure}[t]
    \centering
    \subfigure[Gowalla]{\includegraphics[width=0.49\columnwidth]{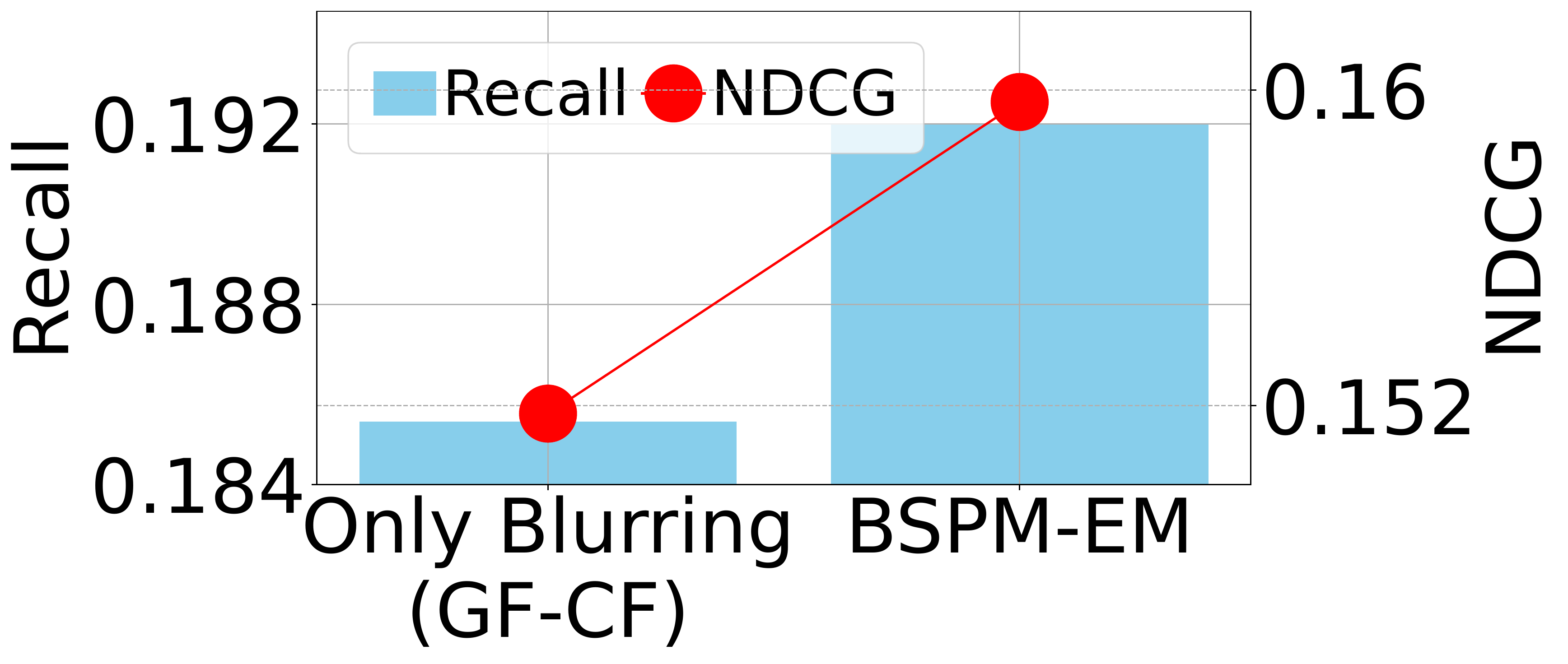}}\hfill
    \subfigure[Gowalla]{\includegraphics[width=0.49\columnwidth]{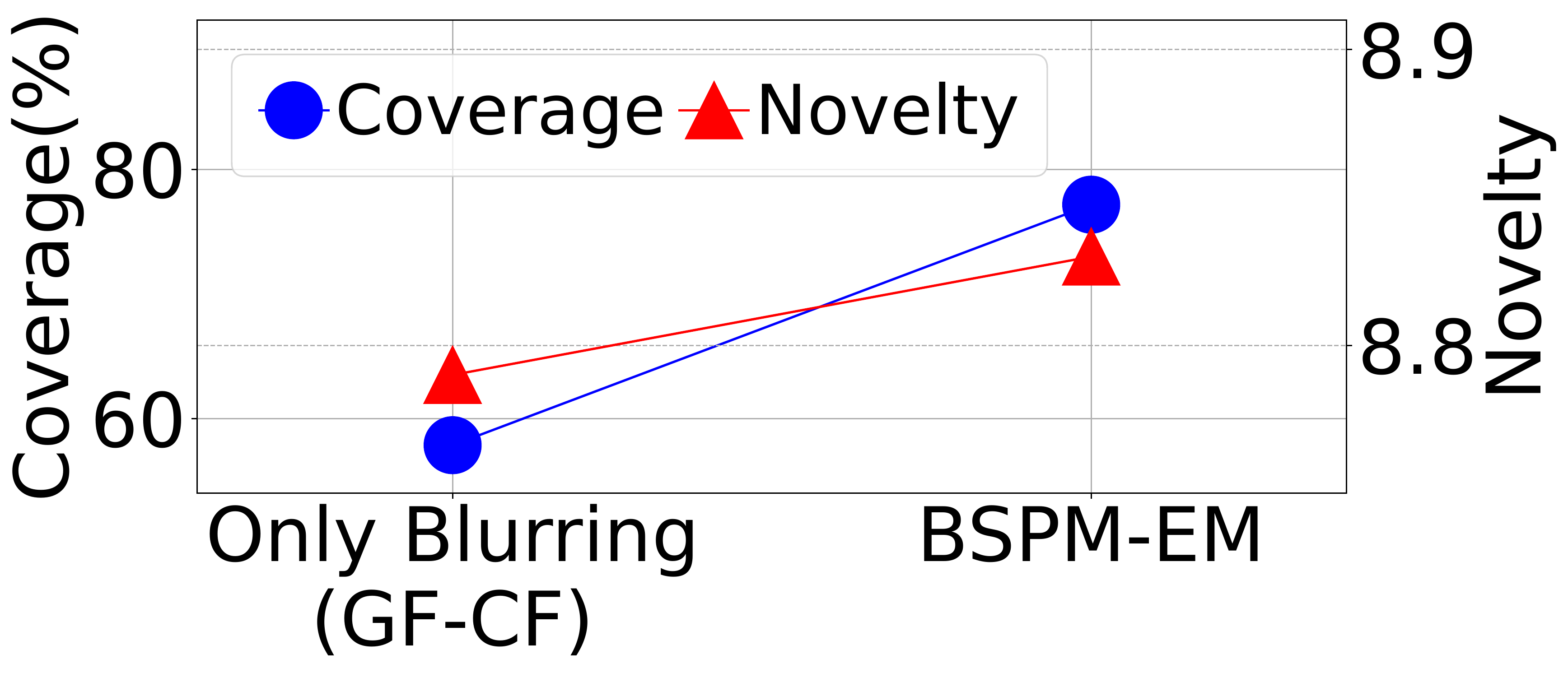}}\\
    \subfigure[Yelp2018]{\includegraphics[width=0.49\columnwidth]{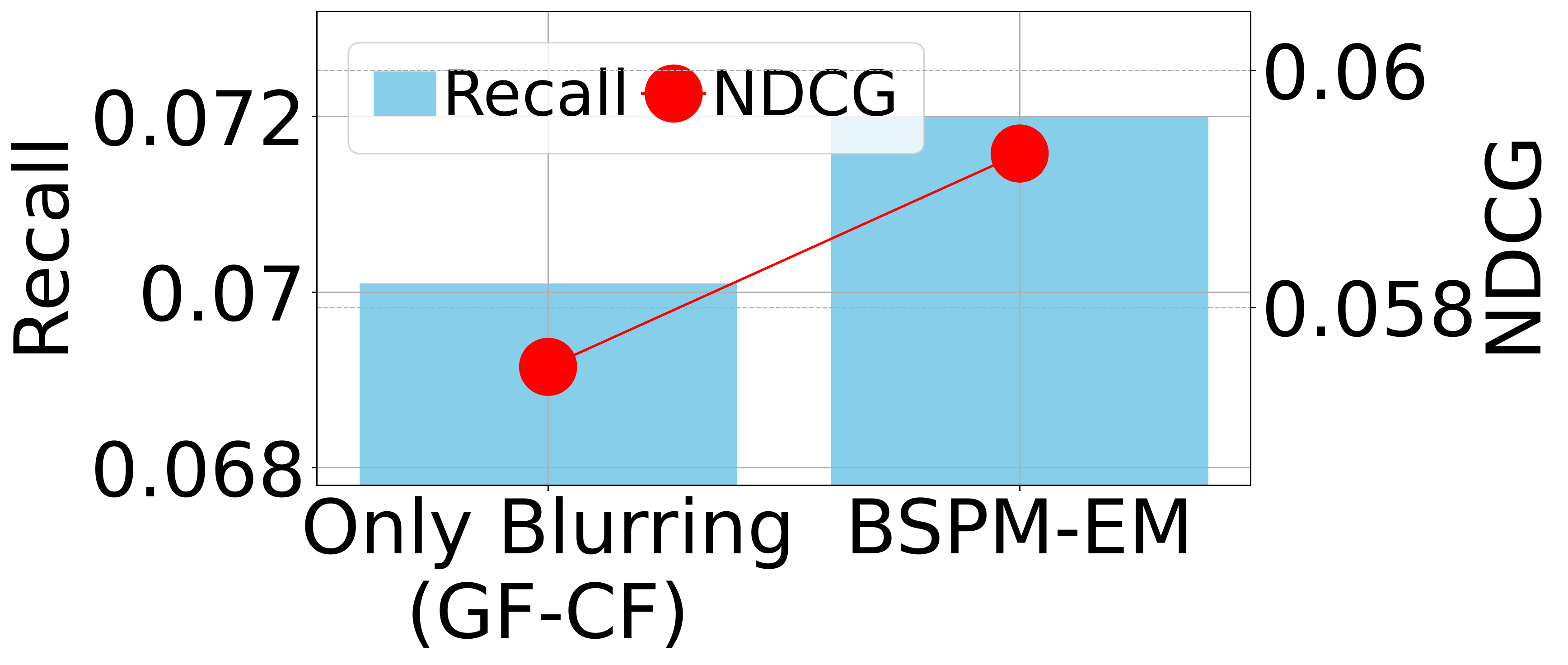}}\hfill
    \subfigure[Yelp2018]{\includegraphics[width=0.49\columnwidth]{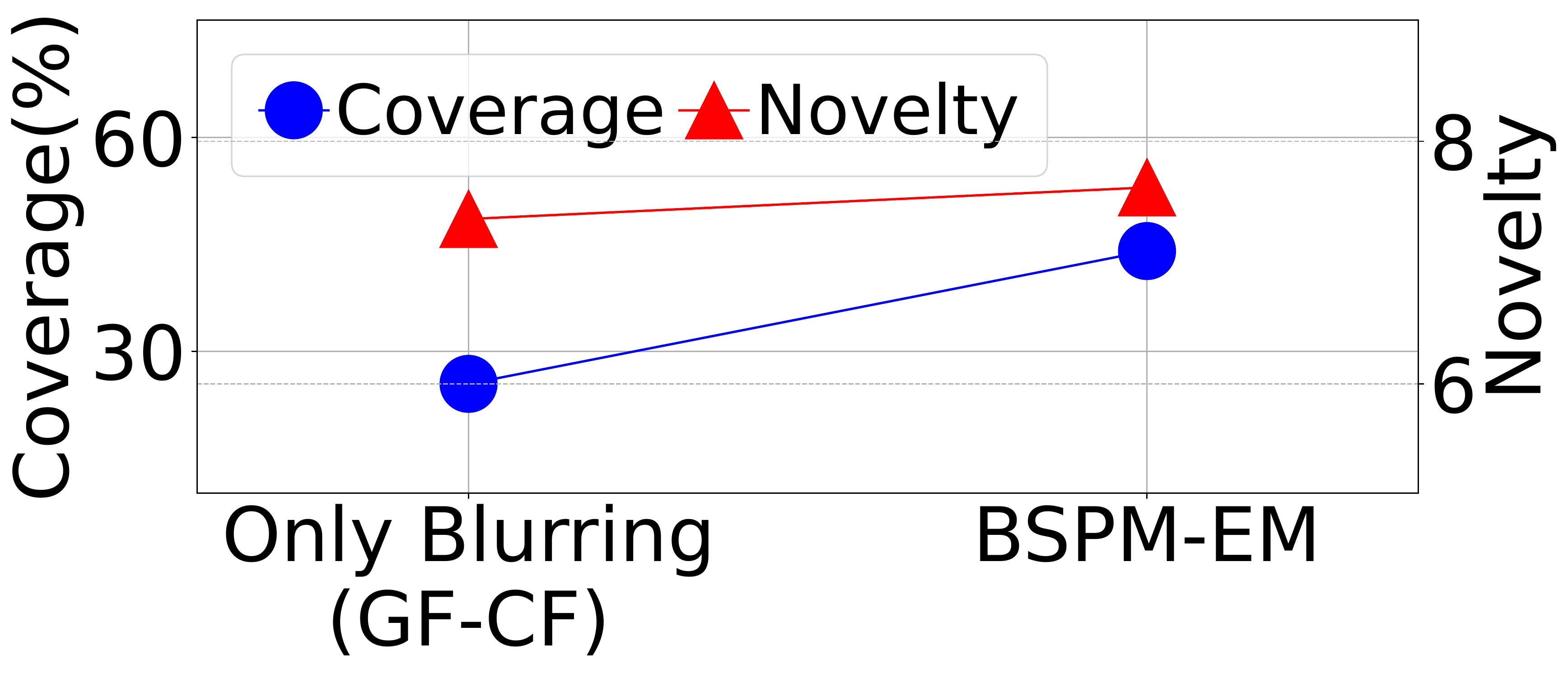}}\\
    \subfigure[Amazon-book]{\includegraphics[width=0.49\columnwidth]{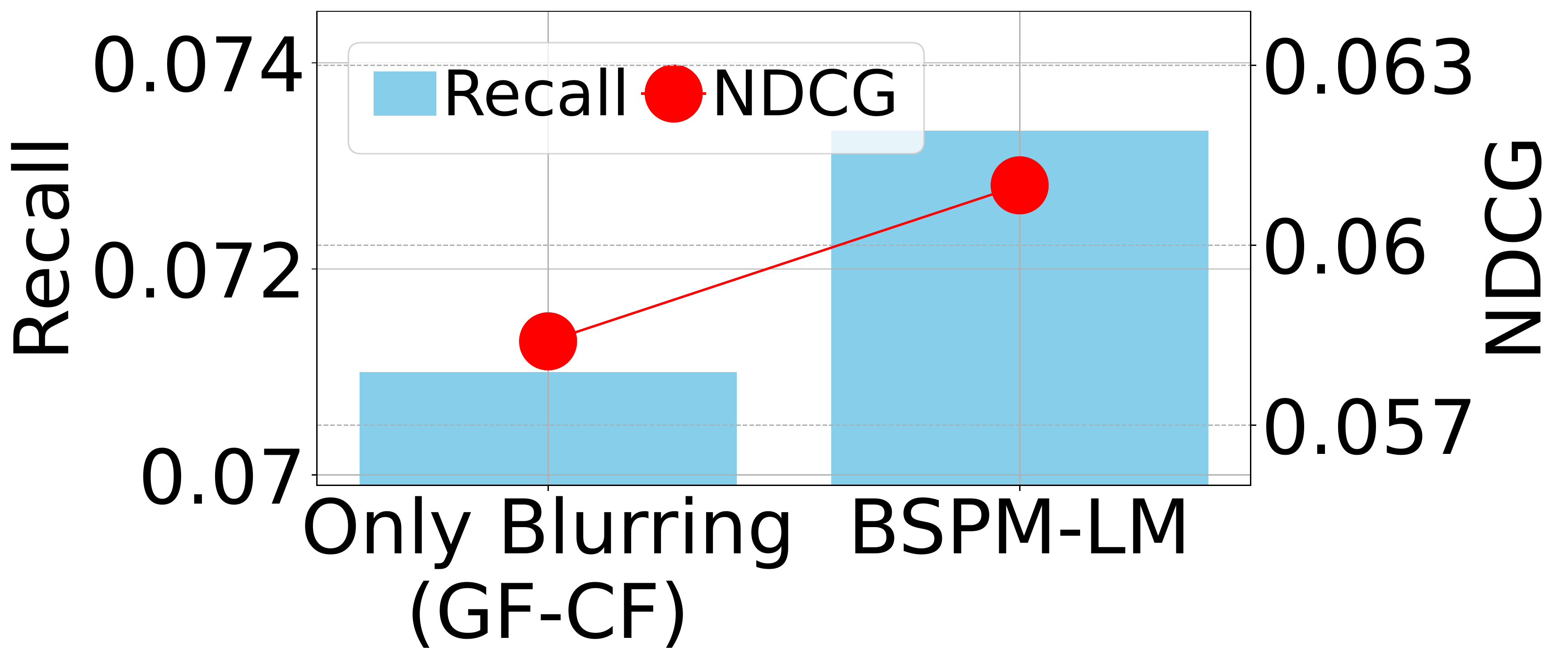}}\hfill
    \subfigure[Amazon-book]{\includegraphics[width=0.49\columnwidth]{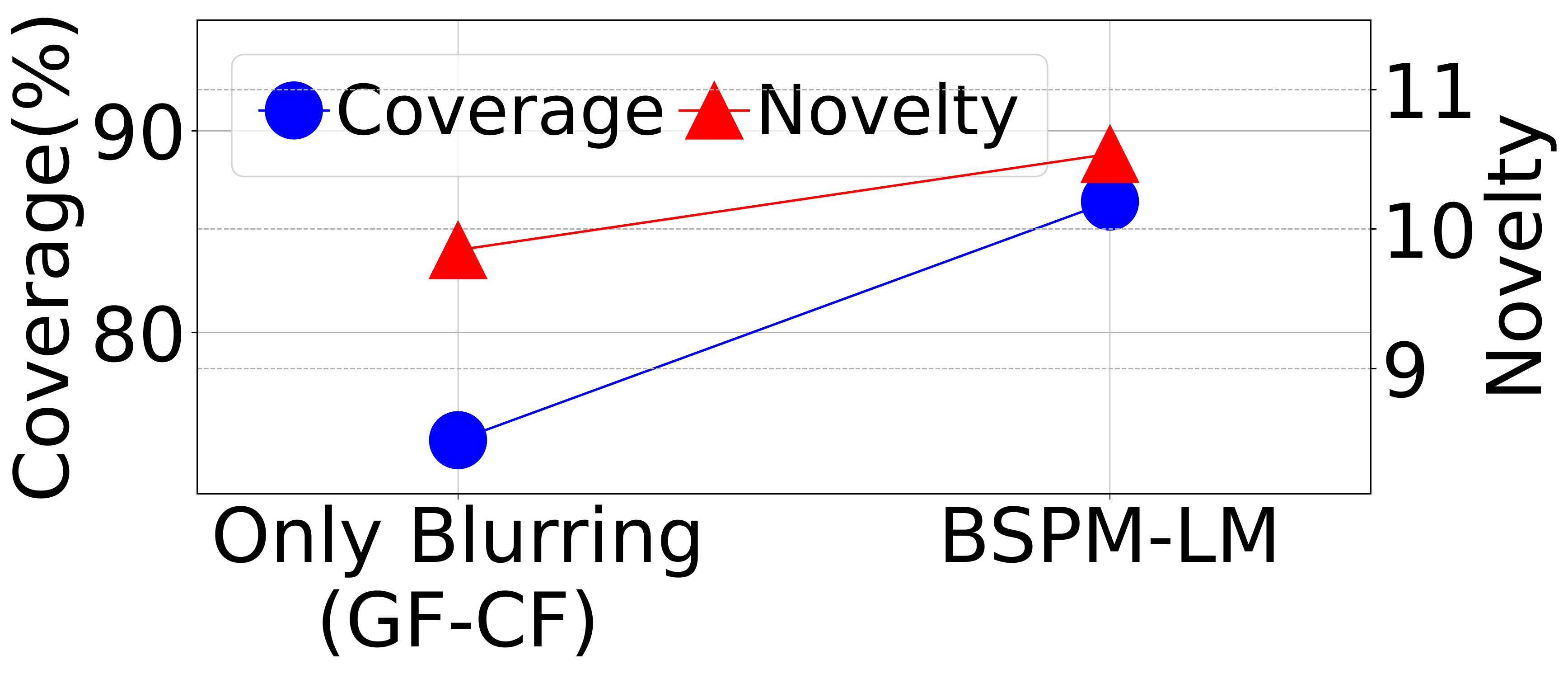}}
    \caption{Accuracy and beyond-accuracy metrics comparison of Only Blurring (GF-CF) and BSPM}
    \label{fig:compare}
\end{figure}

\subsubsection{Ablation study} As ablation study models, we test the models with only a blurring process, denoted `Only Blurring (HE)', `Only Blurring (IDL)', and `Only Blurring (GF-CF)' in Table~\ref{tbl:main_exp}, depending on the used blurring function type. As shown, they do not show reliable performance in comparison with our main model.

We also test whether the residual connection is helpful in each dataset. As reported in Fig~\ref{fig:combi}, it increases the accuracy in Yelp2018. For other datasets, it is best not to use the residual connection.

\subsection{Efficacy of the Sharpening Process}
As mentioned earlier, we are the first proposing to use the sharpening process for CF. Therefore, we conduct in-depth analyses on how the sharpening process contributes. Other metrics for CF to measure the quality of recommendation, called \emph{beyond-accuracy metrics}. For these analyses, we use the following beyond-accuracy metrics: novelty~\cite{zhou2010solving} and item coverage~\cite{Herlocker2004Coverage}. The item coverage refers to the extent of items a recommender system can predict. A novel item for a user is one the user has little or no knowledge about it~\cite{Ge2010beyond}. The novelty measures the unexpectedness of recommended items relative to their global popularity. Using these metrics provides a broader picture of the contribution by the sharpening process. The results are presented in Figs.~\ref{fig:compare} (b), (d), (f), the blurring-sharpening process has higher novelty and coverage scores than the blurring process in all datasets.

It's worth mentioning that coverage increased by 73.20\% on Yelp2018, and the sharpening process improves the ability of the method to recommend long-tail items that users purchase relatively infrequently. In the case of novelty, it can be seen that the metric is improved by 7.01\% compared to that using only the blurring process in Amazon-book. In Figs.~\ref{fig:compare} (a), (c), (e), we can see that the sharpening process is able to enhance Recall and NDCG.

\begin{table}[t]
    \small
    \setlength{\tabcolsep}{3pt}
    \centering
    \caption{The training time of LightGCN, and the pre-processing time of our method. Since our method does not have any training step, we compare our pre-processing time with LightGCN's training time. GF-CF also requires the same types of pre-processing and therefore has the same pre-processing time as ours.}
    \begin{tabular}{c c c c}\toprule
        Model  & Gowalla & Yelp2018 & Amazon-book \\
        \midrule
        Training of LightGCN    & $1.0\times10^4$s & $1.5\times10^4$s & $9.7\times10^4$s \\
        Training of LT-OCF      & $3.6\times10^4$s & $4.8\times10^4$s & $2.3\times10^5$s \\
        Pre-processing of BSPM  & 35.4s & 42.3s & 101.6s\\
        \bottomrule
    \end{tabular}
    \label{tbl:runtime1}
\end{table}

\begin{table}[t]
    \small
    \centering
    \caption{The inference time of BSPM and GF-CF with a mini-batch size of 2048 users}
    \begin{tabular}{c c c c}\toprule
        Model  & Gowalla & Yelp2018 & Amazon-book \\
        \midrule
        GF-CF                   &   9.8s & 10.6s & 40.1s\\
        BSPM-EM (Euler)         &  10.5s & 11.0s & 53.0s\\
        BSPM-EM (RK4)           &  17.9s & 19.8s & 125.5s\\
        BSPM-LM (Euler)         &  10.6s & 11.2s & 59.3s\\
        BSPM-LM (RK4)           &  17.6s & 19.6s & 127.2s\\        
        \bottomrule
    \end{tabular}
    \label{tbl:runtime2}
\end{table}

\subsection{Runtime Analyses}
We also report our pre-processing and inference time in Tables~\ref{tbl:runtime1} and~\ref{tbl:runtime2}. Since our method does not include any training step, it is much faster than other methods --- however, our method requires a pre-processing step to calculate $\tilde{\bm{R}}$, $\tilde{\bm{P}}$, and so on. Among various baselines, LightGCN is one of the simplest and most influential graph convolutional methods, but its training time is several orders of magnitude worse than our pre-processing time. GF-CF also does not have any training step and its pre-processing time is the same as our method. However, our model is more complicated than GF-CF (cf. Eq.~\eqref{eq:gfcf2} for GF-CF vs. Eq.~\eqref{eq:bspm1} or~\eqref{eq:bspm2} for our method) and has longer inference times. However, current official ODE solver implementations on \textsc{PyTorch} do not support sparse matrices. Our method will become much faster with sparse matrix computations.

\subsection{Case Studies}\label{sec:case}
\begin{figure}[t]
    \centering
    \subfigure[User A]{\includegraphics[width=0.48\columnwidth]{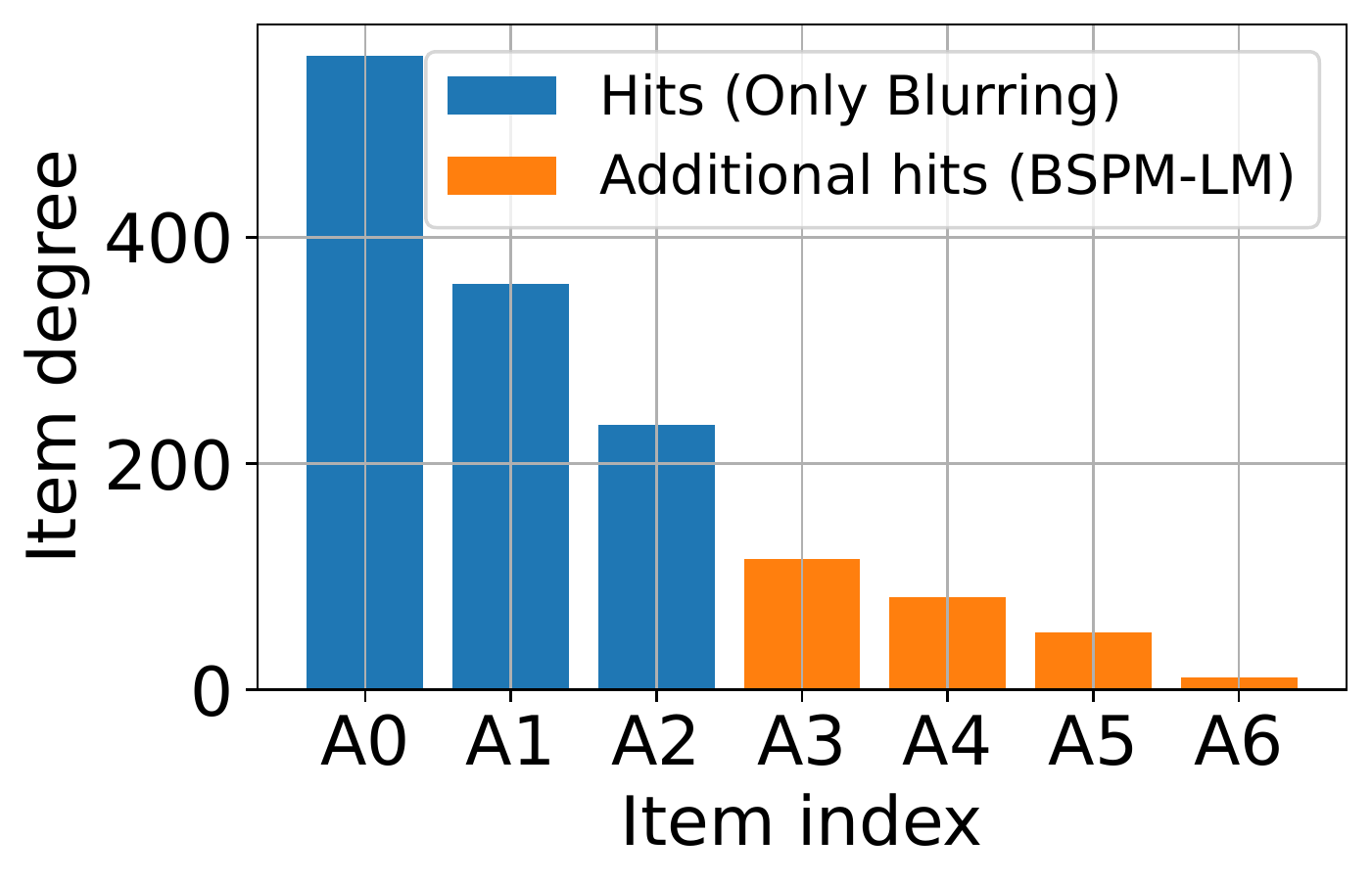}}
    \subfigure[User B]{\includegraphics[width=0.48\columnwidth]{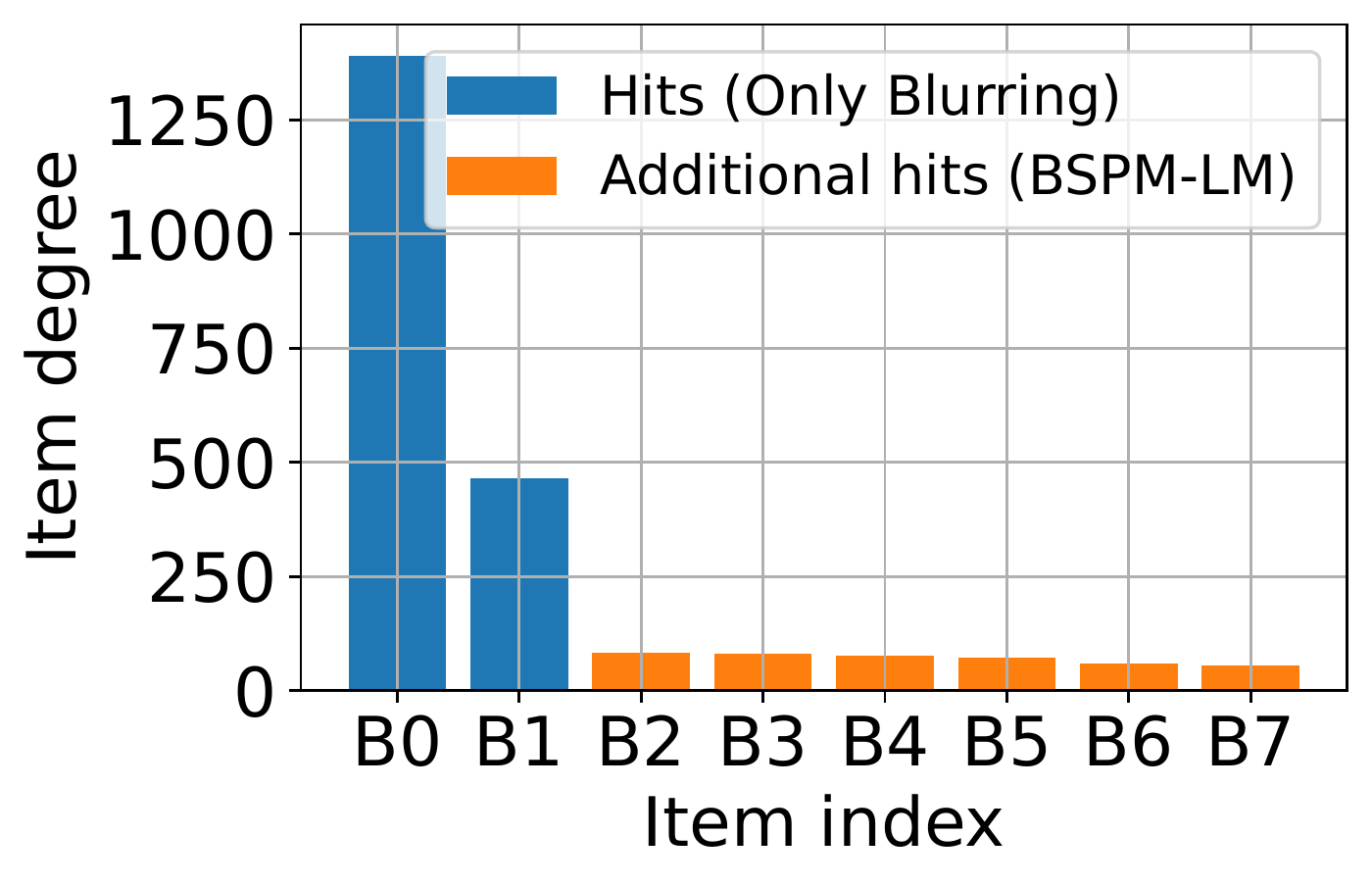}}
    \caption{Node degrees of hits (correctly recommended items) by BSPM-LM on Amazon-book. Blue denotes hits when the sharpening process is skipped. Orange means additional hits after the sharpening process is performed.}
    \label{fig:case_study}
\end{figure}

We compare BSPM-EM and BSPM-LM with and without the sharpening process in terms of Hits@20 to see how the sharpening process changes recommended items. After the sharpening process, more items are accurately recommended for all datasets, including Gowalla, Yelp2018, and Amazon-book. Fig.~\ref{fig:case_study} shows case studies for several users on Amazon-book. After the sharpening process, a few more items are accurately recommended for users A and B.

\begin{figure}[t]
    \centering
    \includegraphics[width=0.55\columnwidth]{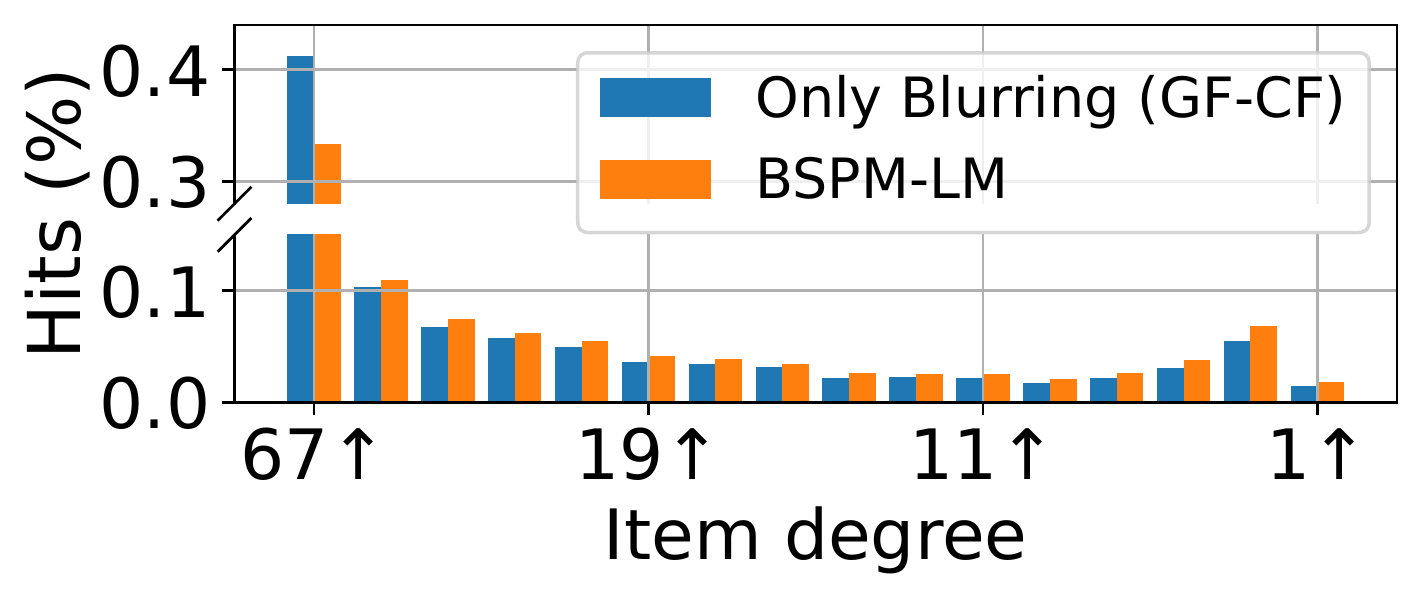}
    \caption{Histogram of Amazon-book's node degrees of correctly recommended items (i.e., hits). In the Y-axis, the number of hits for each node degree bin is divided by the total number of hits.}
    \label{fig:case_study_all}
\end{figure}

Interestingly, those additional hits, highlighted in orange in Fig.~\ref{fig:case_study}, after the sharpening process has relatively low node degrees. In Fig.~\ref{fig:case_study} (b), the recommended items' degrees are high when only the blurring process is performed. In contrast, the degrees of the extra hits after the sharpening process are as low as less than 85, i.e., those additional hits are not popular items but specific to the user. On Amazon-book, the average node degree of hits is 48.20 when only the blurring process is used, but it drops to 33.10 when the sharpening process is added. Fig.~\ref{fig:case_study_all} shows that the ratio of hits with low item degrees increases after the sharpening process. These results suggest that the model with the sharpening process accurately recommends less popular but user-specific items.

The long-tail problem, i.e., how to recommend more user-specific items, is a major challenge in the recommender system. Benchmark data also has long-tail characteristics, as shown in Fig.~\ref{fig:longtail}, making it difficult to recommend user-specific items. Our case study results show that the sharpening process allows for user-specific item recommendations while also improving accuracy.

\section{Conclusion \& Future Work}
We presented a novel paradigm of blurring-sharpening process model (BSPM) for CF. Our work is greatly inspired by the recent two research breakthroughs: i) graph filtering-based methods, e.g., GF-CF, and ii) SGMs for generating fake images. As in SGMs, we adopt the perturbation-recovery paradigm to discover new information. As in GF-CF, we do not learn embedding vectors but directly process the interaction matrix. After defining our BSPM paradigm, we also design a couple of variants to enhance the recommendation accuracy further. In addition, our BSPM is one of the fastest methods ever designed for CF since it does not include any training phase but directly infers unknown user-item interactions. In our experiments with 43 baselines and 3 benchmark datasets, our method marks the best accuracy by large margins. Since our method is not only the most accurate but also one of the fastest methods, it has a significant impact on real-world CF applications.

In the future, we hope that it can be further improved by adopting better blurring and sharpening processes since we have focused on designing the overall architecture by customizing popular filters. In particular, we think that it is promising to learn optimal blurring and sharpening processes from data.

\begin{acks}
Noseong Park is the corresponding author. This work was supported by an IITP grant funded by the Korean government (MSIT) (No.2020-0-01361, Artificial Intelligence Graduate School Program (Yonsei University)) and an ETRI grant funded by the Korean government (23ZS1100, Core Technology Research for Self-Improving Integrated Artificial Intelligence System).
\end{acks}
\bibliographystyle{ACM-Reference-Format}
\bibliography{ref}
    
\end{document}